\documentclass[aps,pra,twocolumn,reprint,superscriptaddress,floatfix]{revtex4-2}

\usepackage{bm,bbm,amsmath,bbold,amssymb,amsbsy,txfonts,graphicx,color,hyperref,xfrac}
\usepackage{textgreek}
\usepackage{CJK}

\newcommand{\kebra}[2]{\vert{#1}\rangle\langle{#2}\vert}
\newcommand{\tr}[1]{\text{Tr}[#1]}
\newcommand{\ket}[1]{|#1\rangle}
\newcommand{\bra}[1]{\langle#1|}
\newcommand{\av}[1]{\langle#1\rangle}
\newcommand{\sprod}[2]{\langle#1|#2\rangle}
\newcommand{\proj}[1]{\ket{#1}\bra{#1}}
\newcommand{\med}[1]{\langle{#1}\rangle}

\begin{document}
	\title{Single quantum emitter Dicke enhancement}
	\author{Tommaso Tufarelli}
	\email{tommaso.tufarelli@nottingham.ac.uk}
	\affiliation{School of Mathematical Sciences and Centre for the Mathematics and Theoretical Physics of Quantum Non-Equilibrium Systems, University of Nottingham, Nottingham, NG7 2RD, United Kingdom}
	\author{Daniel Friedrich}
		\affiliation{Nano-Optics and Biophotonics Group, Experimentelle Physik 5, Physikalisches Institut, Universit\"at W\"urzburg,
D-97074 W\"urzburg, Germany}
\author{Heiko Gro\ss}
	\affiliation{Nano-Optics and Biophotonics Group, Experimentelle Physik 5, Physikalisches Institut, Universit\"at W\"urzburg,
D-97074 W\"urzburg, Germany}
	\author{Joachim Hamm}
		\affiliation{The Blackett Laboratory, Department of Physics,
Imperial College London, London SW7 2AZ, United Kingdom}
	\author{Ortwin Hess} 
	\affiliation{The Blackett Laboratory, Department of Physics, Imperial College London, London SW7 2AZ, United Kingdom}
	\affiliation{School of Physics and CRANN Institute, Trinity College Dublin, Dublin 2, Ireland}
	\author{Bert Hecht}
	\email{hecht@physik.uni-wuerzburg.de}
	\affiliation{Nano-Optics and Biophotonics Group, Experimentelle Physik 5, Physikalisches Institut, Universit\"at W\"urzburg,
D-97074 W\"urzburg, Germany}

		\begin{abstract}
		Coupling $N$ identical emitters to the same field mode is well-established method to enhance light matter interaction. However, the resulting $\sqrt{N}$-boost of the coupling strength comes at the cost of a ``linearized" (effectively semi-classical) dynamics. Here, we instead demonstrate a new approach for enhancing the coupling constant of a \textit{single} quantum emitter, while retaining the nonlinear character of the light-matter interaction. We consider a single quantum emitter with $N$ nearly degenerate transitions that are collectively coupled to the same field mode. We show that in such conditions an effective Jaynes-Cummings model emerges, with a boosted coupling constant of order $\sqrt N$. The validity and consequences of our general conclusions are analytically demonstrated for the instructive case $N=2$. We further observe that our system can closely match the spectral line shapes and photon autocorrelation functions typical of Jaynes-Cummings physics, hence proving that quantum optical nonlinearities are retained. Our findings match up very well with recent broadband plasmonic nanoresonator strong-coupling experiments and will therefore facilitate the control and detection of single-photon nonlinearities at ambient conditions.
		
	\end{abstract}
	
	\maketitle
	
	\section{Introduction} 	
	\begin{figure*}[t!]
	    \centering
	    \includegraphics[width=.99\linewidth]{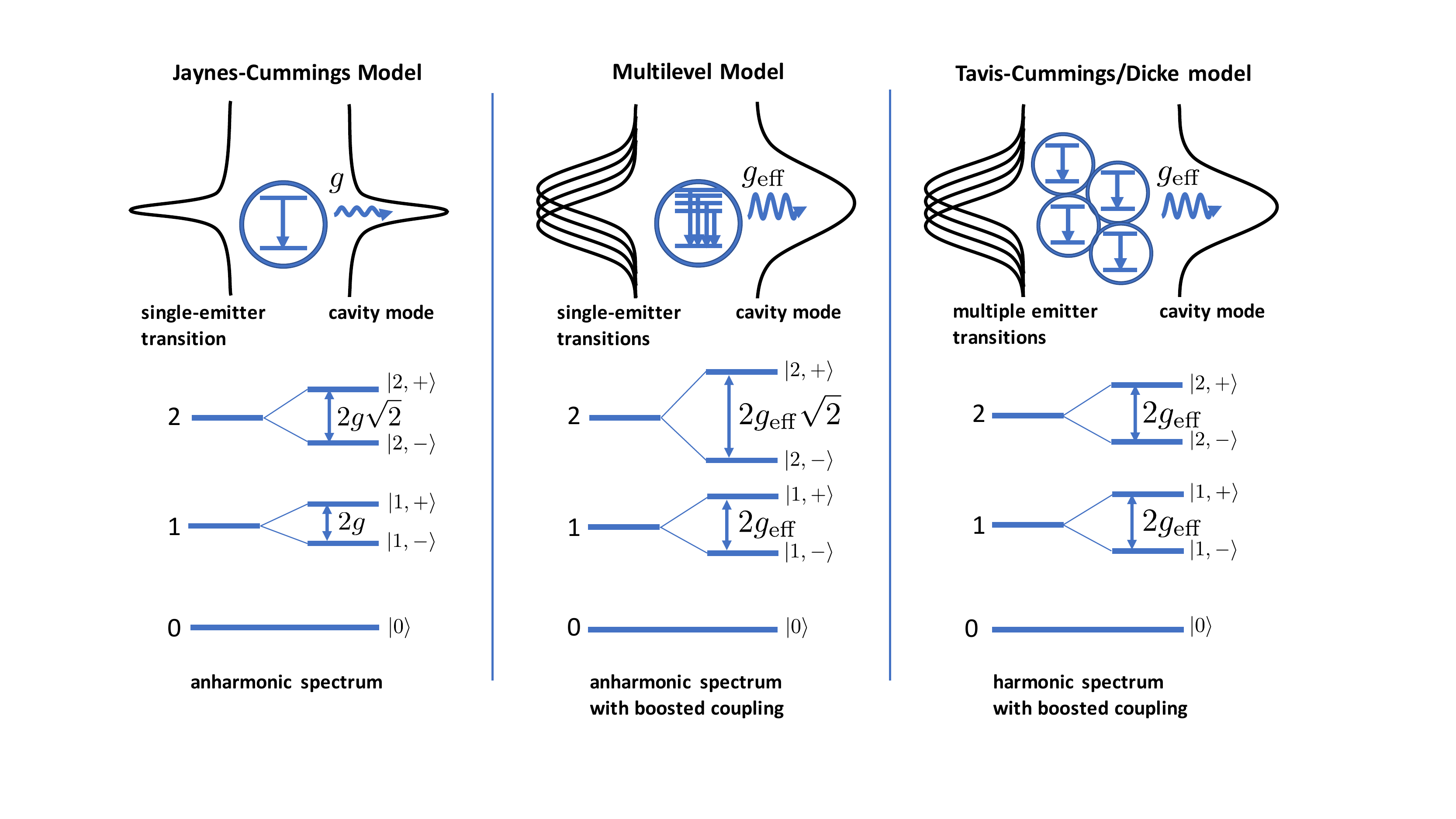}
	    \caption{Visual summary of the light-matter interaction models treated here. For multilevel, TC and Dicke models, only energy levels that have a dirrect correspondence in the JCM are shown. \textbf{Left:} Standard Jaynes-Cummings model, in which a single two-level emitter couples resonantly to a single mode field, with strength $g$. The resulting anharmonic spectrum is a cornerstone of modern quantum optics, but challenging to probe experimentally: the required regime $g\gg \kappa$ is difficult to achieve, where $\kappa$ is the cavity decay rate. \textbf{Center:} Multilevel model, the focus of this paper. A quasi-degenerate multilevel emitter, with $N$ closely spaced excited states, interacts with a single mode field. When all the transitions are approximately resonant with the field, an effective coupling constant $g_{\sf eff}$, enhanced by a factor $\propto \sqrt{N}$, is established. In principle, the resulting low energy spectrum can closely approximate the JC spectrum for any value of $N$. \textbf{Right:} Tavis-Cummings and Dicke Models, in which $N\gg1$ nearly-identical two level systems interact with the same field. A boosted effective coupling constant $g_{\sf eff}\propto \sqrt{N}$ is established also in this case, but the spectrum becomes approximately harmonic as $N$ is increased.}
	    \label{Fig-summary}
	\end{figure*}
	Strongly coupled light-matter systems, described by the Jaynes-Cummings (JC) model \cite{shore1993jaynes}, are of fundamental interest in testing the quantised nature of coupled light and matter degrees of freedom in the context of cavity QED \cite{haroche2006exploring}. Importantly, JC physics requires that a \textit{single} two-level emitter interacts with a single-mode electromagnetic field. Such systems can be relevant for quantum information processing, for example in interfacing flying and static qubits \cite{reiserer_quantum_2014} or for quantum-state engineering~\cite{haroche2006exploring}. Yet strong coupling of light and a single emitter is not easy to achieve in practice, due to the large mismatch in the spatial extension of free-space photons and typical emitters (i.e. artificial or real atoms) that feature electronic transitions in the optical domain \cite{Biagioni,walther2006cavity}. 
	
	A possible route to overcome this difficulty is to engineer an ``effective" emitter with large dipole moment. This is most commonly implemented by combining $N\gg1$ identical emitters coupled to the same field, resulting in the well-known ``Dicke enhancement" of light-matter interaction: the effective coupling strength is boosted by a factor of order $\sqrt N,$ compared to the single-emitter case \cite{torma_strong_2015,dovzhenko_lightmatter_2018,yu_strong_2019,g-aggregates}. However, the relevant theory in such systems is the many-emitter limit of the Tavis-Cummings (TC) model \cite{tavis1968exact}, whose low-energy dynamics is effectively \textit{linear,} i.e., indistinguishable from that of coupled harmonic oscillators \cite{Quesada-2012}.
	
\textit{The key achievement of our work is to bypass this seemingly inevitable trade-off between coupling strength and (quantum) nonlinearity: we open an innovative route to a `Dicke-like' enhancement of the coupling constant which preserves the precious quantum optical nonlinearities of the JC model.}

In this paper we indeed show how the exploitation of a multilevel emitter, with $N$ nearly degenerate excited states coupled to the same field mode, increases the effective light-matter coupling constant by a factor of oredr $\sqrt{N}$. While this is the same scaling found in the Dicke \cite{DickeOriginal} and TC models \cite{tavis1968exact} with $N$ emitters, the crucial difference is that in our case the quantum non-linearity of the interaction is preserved: we demonstrate this based on emission spectra and second-order photon autocorrelation functions. More precisely, our system can closely approximate the behaviour of a JC model, in principle, for arbitrary values of $N$. These findings, in their simplest form, are schematised in Fig~\ref{Fig-summary}. 

	Additionally, we find that our predictions are robust against typical sources of decoherence (dephasing) that are common at elevated temperatures. Thus, while our results are fully general and independent of the physical implementation of the model, we anticipate that our approach can be particularly useful in the context of \textit{room temperature} strong coupling. We thus find it useful to provide some context on the current state-of-the-art of this research field, and to highlight how our research can help it move forward.
	
	Recently, room-temperature strong coupling with many emitters~\cite{molecular,baranov_ultrastrong_2020,Geisler-2019,moilanen_active_2018} and even single quantum emitters \cite{Santhosh,Baumberg,our_paper,leng_strong_2018,park_tip-enhanced_2019} have been experimentally demonstrated. These experiments rely on the remarkable coupling strengths achievable in plasmonic nanoresonators, thanks to the deeply sub-wavelength mode volumes of the latter. In order to relax the mode volume requirements (which are currently pushing the limits of nanofabrication techniques) and obtain more robust experimental realizations, it would be of great interest to find emitter systems with increased dipole moments. This would result in larger coupling constants while still allowing to harness the single-emitter quantum nonlinearity of the JC model (JCM) \cite{Kasprzak-2010, Fink-2008}. So far, only the ``first rung" of the JC ladder, i.e. the single-excitation regime of the model, has been experimentally detected at ambient conditions. Efforts are still ongoing to detect higher rungs of the ladder to shed light on (and harness) the quantum non-linearity of the model.
	
	In order to realise the JCM experimentally, great efforts have been made over the years in the fabrication of artificial atoms --- such as semiconductor quantum dots~\cite{Reithmaier} --- that are close to ideal two-level emitters. Indeed, so far the multi-level nature of such emitters has mainly been treated as a modelling nuisance, i.e. as a source of errors with respect to an idealised two-level scenario. In this context, Madsen et al.~\cite{Madsen} and our previous paper~\cite{our_paper}, have recently mentioned the possible effects of a more complicated energy level structure of the emitter. To the best of our knowledge, however, our previous work~\cite{our_paper} was the first to explicitly suggest that an emitter with multiple ($>2$) energy levels can even bring about some advantages: in the right conditions it may be used to simulate a JC system with \textit{enhanced} coupling constant. With reports of quantum dots featuring as much as 64 quasi-degenerate excited states \cite{Liu-2010,Hens-2012}, a general quantum-mechanical analysis of such a system (\textit{multilevel model} for brevity) is in order. 

	The findings presented here will thus be of importance for the selection and the design of suitable emitters for nanoscale strong-coupling experiments at ambient conditions and for the interpretation of such experiments. Importantly, our approach lives independently of the contentious issues regarding the calculation of mode volumes and coupling constants in quantum (and classical) nanophotonics, see for example~\cite{NV_review,Nuttawut_quasimodes,Hughes_quasimodes}. These approaches will of course determine the numerical values of coupling constants in different experimental situations, but they do not affect the nature of our quantum optical model and the general conclusions we can draw from it.

	The paper is organised as follows. In Section~\ref{modelsection} we introduce a Hamiltonian that models an emitter with $N$ excited levels coupled to the same ground state via a single mode field. We will show that our system can be understood as a JC model which is weakly interacting with $N-1$ ``dark states" of the emitter. In Section~\ref{MEQsec} we generalise our model to an open system via a master equation. This takes into account incoherent processes such as photon loss, dephasing and emitter pumping, all of which may be required for describing realistic experimental scenarios. In Section~\ref{2sublevels} we support our general arguments with detailed analytical results for the special case $N=2$: we find that our system's emission spectrum features approximately the same resonances as a JC model, plus additional features (due to the dark states) that we fully characterize. Section \ref{numericalsection} presents numerical studies for $N=3$, comparing common experimental observables in our model with the same quantities calculated in reference JC and TC/Dicke models.
	In Section~\ref{conclusions} we draw our conclusions. 
	For completeness, Appendix~\ref{auxmodels} contains some useful reminders and formulas pertaining JC, TC and coupled-oscillator models, since those are referenced throughout the paper.

	\section{Model}\label{modelsection}
	
	Our \textit{multilevel model} features a quantum emitter with electronic ground state $\ket G$ and a collection of excited states $\ket{e_k}$, also called \textit{sublevels}, where $k=1,2,\dots,N$. Each internal transition of the emitter, $\ket G\leftrightarrow \ket{e_k}$, is coupled to a single-mode field (cavity for brevity) with strength $g_k\in\mathbb{R}$. The typical physical system we have in mind is a multilevel quantum dot embedded in a plasmonic nanoresonator, where the cavity resonance is \textit{broadband}, in the sense that it overlaps with all the emitter transitions, and where strong field gradients may allow for light-matter couplings beyond the dipole approximation (hence the possibility of coupling a large number of levels). Yet, it is important to note that our quantum optical formalism applies to any type of emitter or resonator that satisfy our modelling assumptions. As usual the cavity mode is described via the annihilation operator $\hat a$, with $[\hat a,\hat a^\dagger]=1$.

	Taking $\hbar=1$ for convenience, the cavity resonance has average frequency (hence energy) $\omega_0$, the energy of $\ket G$ can be set to zero without loss of generality, while each state $\ket{e_k}$ has energy $\omega_0+\Delta_k$. Here $\Delta_k$ is the detuning between the $k$-th emitter transition ($\ket{e_k}\to\ket G$) and the cavity. For convenience we assume $\Delta_1\le\Delta_2\le...\le\Delta_N$, we indicate the full range of energy detunings as $\varepsilon\equiv\Delta_N-\Delta_1$ and the average detuning as $\Delta=\tfrac{1}{N}\sum_{k}\Delta_k$.  we can write the total Hamiltonian as
	\begin{align}
	H&=H_0+V,\label{Hmany}\\
	H_0&=\omega_0\hat a^\dagger \hat a+\sum_{k=1}^n(\omega_0+\Delta_k)\proj{e_k},\\
	V&=\sum_{k=1}^{N}g_k\left(\hat a\,\kebra{e_k}{G}+\hat a^\dagger\kebra{G}{e_k}\right),
	\end{align}
	where $H_0,V$ are, respectively, the free and interaction Hamiltonians, and we have neglected counter-rotating terms in the light-matter interaction. The physics of Hamiltonian~\eqref{Hmany} can in general be quite complicated, despite $H$ being number-conserving --- i.e. $H$ commutes with the number operator $\hat{\mathcal N}_{\mathrm{{tot}}}=\hat a^\dagger\hat a+\sum_k\proj{e_k}$. However, here we are interested in the special case of quasi-degenerate emitters that are near-resonant with the cavity, i.e. we assume the parameter regime $|\Delta_k|\ll g_k$. This approximation holds e.g. for colloidal quantum dots coupled to plasmonic nanoresonators. Here, $\Delta_k$ are on the order of a few meV, while the coupling strengths, $g_k$, are on the order of 100meV \cite{our_paper}. In this scenario a number of qualitative observations can be made. 
	\begin{figure}[t!]
		\includegraphics[width=\linewidth]{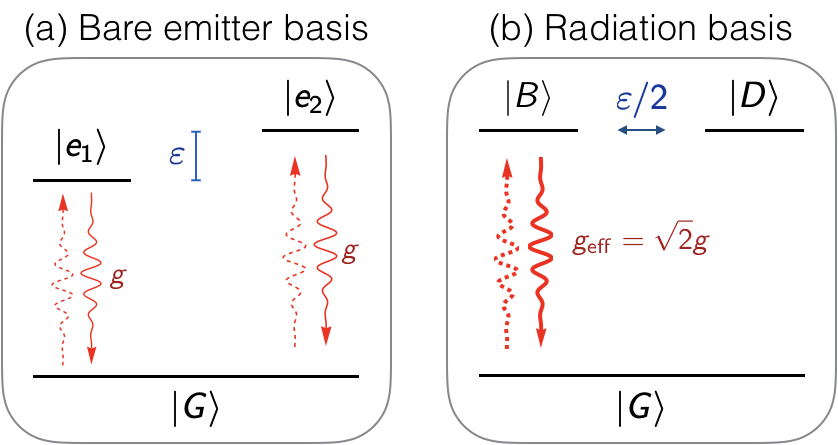}
		\caption{Illustration of the transformation between the bare emitter basis and the radiation basis for $N=2$ sublevels, and the resulting appearance of a boosted effective coupling. The same principles hold for larger values of $N$. \textbf{(a)} In the bare emitter basis the two levels $\ket{e_{1,2}}$ are not directly coupled to each other, while $\varepsilon$ plays the role of a relative detuning between them. Both $\ket{G}\leftrightarrow \ket{e_{1,2}}$ transitions couple to the cavity field, and we are assuming for simplicity that the associated coupling constants are the same: $g_1=g_2=g$. \textbf{(b)} In the radiation basis, the bright and dark states $\ket B, \ket D$ have the same frequency, but are coupled to each other with strength $\varepsilon/2$. The transition $\ket G\leftrightarrow \ket{B}$ is coupled to the field with enhanced strength $g_\mathrm{eff}=\sqrt2 g$, while the dark state $\ket D$ does not directly interact with the field.}\label{Figlivelli}
	\end{figure}
	\subsection{Bare emitter basis versus radiation basis}
	To aid understanding, the main ideas behind this subsection are illustrated in Fig.~\ref{Figlivelli} for the simplest case, $N=2$. We begin our study of the general $N$-sublevel model by considering the case of exact degeneracy of the excited states, $\Delta_k=\Delta\,\forall k$, or equivalently $\varepsilon=0$ in terms of the full range of energy detunings. The resulting mathematics follow in a straightforward manner from the well-known JCM. This simple starting point will set the stage to investigate small deviations from an idealized scenario, i.e. cases with small but nonzero $\varepsilon$. For $\varepsilon=0$, we find that any linear combination of the states $\ket{e_k}$ is itself a valid emitter excited state, or more precisely, an eigenstate of $H_0$ with eigenvalue $\omega_0+\Delta$. Among all the states that can be constructed in this way, of particular significance is the coherent superposition
	\begin{equation}
	\ket{B}=\frac{\sum_k g_k\,\ket{e_k}}{\sqrt{\sum_{j}g_{j}^2}},\label{bright}
	\end{equation}
	where the denominator ensures normalization. We shall refer to $\ket{B}$ as the  \textit{bright state}, or \textit{superradiant state} (mimicking the terminology of the Dicke model). State $\ket B$ can be interpreted as the only emitter excited state that couples directly to the cavity field. Indeed it is easy to check that the light-matter interaction term in Eq.~\eqref{Hmany} may be rewritten in the simple form
	\begin{align}
	V&=g_\mathrm{eff}\left(\hat a\,\kebra{B}{G}+\hat a^\dagger\kebra{G}{B}\right),\label{Vnew}
	\end{align}
	where we have defined an effective coupling constant
	\begin{equation}
	g_\mathrm{eff}=\sqrt{\sum_{k}g_{k}^2}.\label{geff}
	\end{equation}
	Hence, through an appropriate change of basis for the emitter, the $N$ degenerate transitions $\ket G\leftrightarrow \ket{e_k}$ can be replaced by a single transition $\ket G\leftrightarrow \ket{B}$, characterized by an enhanced coupling to the cavity field. Eq.~\eqref{geff} indeed shows that the effective coupling scales like $g_\mathrm{eff}\sim\sqrt N$ with the number of sublevels. This boosting of the effective light-matter coupling is analogous to what happens in the Dicke \cite{DickeOriginal} and TC \cite{tavis1968exact} models, where $N$ degenerate emitters are coupled to the same field. A crucial difference from the TC model is that Eq.~\eqref{Vnew} retains the structure of a JC interaction Hamiltonian, so that in the model studied here there is in principle no loss of anharmonicity as the number of sublevels $N$ is increased.
	
	To fully specify the mentioned change of basis, we can define $N-1$ further linear combinations of the excited states $\ket{e_k}$, say $\{\ket{D_1},\ket{D_2},\dots,\ket{D_{N-1}}\}$, which together with $\{\ket{G},\ket{B}\}$ form a complete orthonormal set for the emitter internal levels. When $\varepsilon=0$ each $\ket{D_k}$ is also an eigenstate of $H_0$ with eigenvalue $\omega_0+\Delta$, due to the $N$-fold degeneracy. By construction, these additional basis states \textit{do not} directly interact with the cavity field, i.e. 
	\begin{equation}
	V\ket{D_k}=\bra{D_k}V=0,\qquad k=1,2,\dots,N-1,\label{nointeraction}
	\end{equation}
	and may be called \textit{subradiant states} or \textit{dark states}. In practice, explicit expressions for the dark states may be found by standard orthogonalization procedures (details not shown). The above discussion suggests that the set $\{\ket{G},\ket{B},\ket{D_1},\ket{D_2},\dots,\ket{D_{N-1}}\}$, which we will call \textit{radiation basis}, may be a more meaningful conceptual tool to study light-matter interaction in our system, as compared to the \textit{bare emitter basis} $\{\ket{G},\ket{e_1},\ket{e_2},\dots,\ket{e_N}\}$. 
	
	For the purpose of modelling the optical properties of the system, it is now tempting to simply neglect all the dark states and retain only the two emitter levels $\ket{B},\ket G$. This is justified if there are no additional processes, coherent or incoherent, that are able to populate the dark states and/or couple them to the two levels $\ket{B},\ket{G}$. When all these conditions are satisfied, the system is \textit{exactly} described by a JCM with coupling constant $g_\mathrm{eff}$.\\
	
	We are now in a position to study small deviations from the ideal case. As anticipated, the most obvious deviation from pure JC physics occurs when the excited states $\ket{e_k}$ are not exactly resonant with each other (i.e. $\varepsilon\neq0$). In the radiation basis, the (small) relative detunings between the levels $\ket{e_k}$ will translate into (weak) couplings between the bright and dark states, and between the dark states themselves. In other words, $\ket{B}$ and $\ket{D_k}$ cease to be eigenstates of $H_0$ when $\varepsilon\neq0$. Note however that Eqs.~\eqref{Vnew} and \eqref{nointeraction} remain valid in general, so that the dark states may interact with the field only \textit{indirectly}, i.e. through the mediation of state $\ket{B}$. These additional interactions will bring about deviations from the well understood JC physics described above. While a general calculation of these effects would be cumbersome and beyond the scopes of this paper, the case $N=2$ sublevels is simple enough to be treated analytically and already contains all the essential ingredients in our problem --- see Section~\ref{2sublevels} below and Fig.~\ref{Figlivelli}. 
	
	In some cases, the exact equivalence to a JC system may be lost even in the case of perfect degeneracy between the excited sublevels. This can happen if incoherent processes couple the pair $\{\ket{G},\ket{B}\}$ to the dark states. We shall see in Section~\ref{2sublevels}, for example, that emitter dephasing provides a common mechanism for interconverting bright and dark states. Yet, our analytical and numerical results suggest that the multilevel model can provide an excellent approximation to JC physics even when these imperfections are taken into account.
	
	\section{Master Equation}\label{MEQsec}
	\begin{figure}[t!]
		\includegraphics[width=1.05\linewidth]{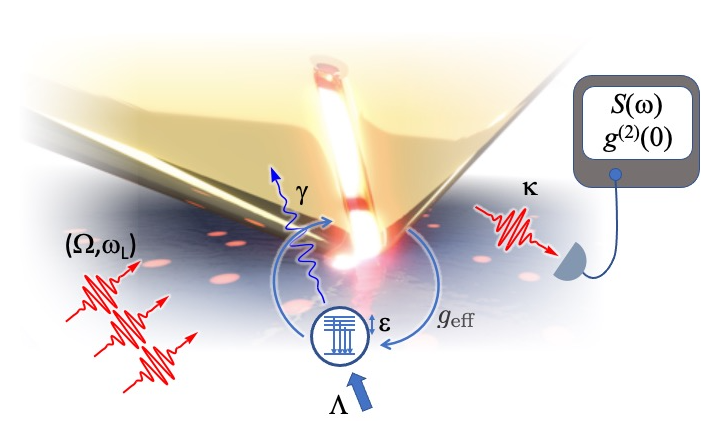}
		\caption{Schematic of the investigated setup. The system may be excited through an incoherent pump of intensity $\Lambda$ applied to the emitter, or coherently driving of the (plasmonic) cavity with field amplitude $\Omega$ and frequency $\omega_L$. The bright state of the emitter is coupled to the cavity with the effective strength $g_\mathrm{eff}$, and the emitter sublevels are spread in frequency over a range $\varepsilon$. $\kappa$ is the loss rate of the cavity and $\gamma$ the decay rate of all the emitter sublevels, which are also subject to dephasing at rate $\gamma_d$ (not shown). All the observables discussed in this work can be measured by collecting the cavity output field.}\label{figaschema}
	\end{figure}
	In order to predict relevant observables for future experiments, we now introduce an open system generalisation of our model, via the well established Gorini-Kossakowski-Sudarshan-Lindblad (GKSL) master equation~\cite{Breuer}. As anticipated above, the main experimental scenario we have in mind is a quantum dot interacting with a plasmonic nanoresonator, yet the ideas presented here may be applicable to a wider variety of setups. We will include cavity photon loss as the primary channel for energy dissipation, while the decay rate of the emitter into free-space modes will be assumed negligible (however, in simulations it will remain finite for reasons of numerical stability). Such an assumption is reasonable due to the large $\beta$-factors achievable in plasmonic resonators. On the other hand, we shall assume that the emitter excited states can be broadened well beyond their radiative linewidth by additional dephasing processes -  typically thought of as a drawback of room temperature operation. We consider two ways in which energy can be supplied to the coupled system: incoherent pumping of the emitter, or a coherent driving field applied to the cavity. While the first is a common method to feed excitations into quantum dots, the second is arguably the standard avenue to test nonlinearities in a quantum optical system. For a visual summary, the main ingredients of our open system model are sketched in Fig.~\ref{figaschema}. In formulas, the following master equation will be used to model the emitter-cavity system,
	\begin{equation}
	\dot\rho=-i[H+V_\mathrm{drive},\rho]+\sum_{L\in\mathcal{J}}\frac12\left(2L\rho L^\dagger-L^\dagger L\rho-\rho L^\dagger L\right),\label{master}
	\end{equation}
	where $H$ is the Hamiltonian \eqref{Hmany}, and
	\begin{align}
	V_\mathrm{drive}=\Omega\left(\hat a e^{i\omega_Lt}+\hat a^\dagger e^{-i\omega_Lt}\right)
	\end{align}
	is an interaction term describing coherent driving of the cavity by a classical laser at frequency $\omega_L$ ($\Omega$ quantifies the driving field strength). The $L$'s appearing on the right hand side of Eq.~\eqref{master} are the \textit{jump operators}, used to model incoherent processes. They are collected in the set ${\cal J}$, which features the following elements:
	\begin{itemize}
		\item Cavity decay (photon loss) at rate $\kappa$: $L=\sqrt{\kappa}\,\hat a$
		\item Radiative decay of the emitter $\gamma$: $L=\sqrt{\gamma} \,\kebra{G}{e_k}$,
		\item Emitter dephasing $\gamma_d$: $L=\sqrt{\gamma_d}\,\ket{e_k}\bra{e_k}$,
		\item Emitter pumping of (total) rate $\Lambda$: $L=\sqrt{\frac{\Lambda}{N}} \,\kebra{e_k}{G}$,
	\end{itemize}
	where $k=1,2,\dots,N$. We assumed that the decay and dephasing rates are the same for all sublevels, and that the pump power $\Lambda$ is split symmetrically over the $N$ sublevels. We also consider equal couplings $g_k=g_\mathrm{eff}/\sqrt{N}$ for all $k$, while the detunings $\Delta_k$ are equally spaced in the interval $[\Delta-\varepsilon/2,\Delta+\varepsilon/2].$ All the assumed symmetries between the sublevels are of course an idealisation. Many generalisations of this basic model can be included in principle. However, we do not expect that these can bring about significant qualitative departures from the physics described here. Hence, in this work we confine our analysis to the above setting for simplicity and computational efficiency.

	\section{Analytical results for 2 sub-levels}\label{2sublevels}
\begin{figure}[htp]
		\includegraphics[width=\linewidth]{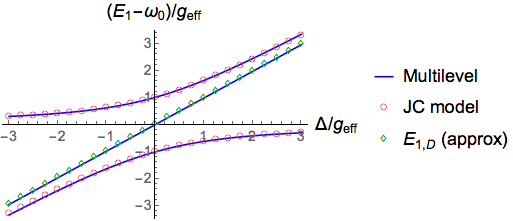}
		\caption{Eigenvalues of $\mathbb H_1$ as a function of the average detuning $\Delta$, for $\varepsilon=0.25g_\mathrm{eff}$ (blue continuous lines). The anticrossing behaviour of eigenvalues $E_{1,\pm}$ around $\Delta=0$ approximates well (the maximum relative error in the plotted range being $\sim1.3\%$) the reference JCM with coupling $g_\mathrm{eff}$ (red circles). The energy $E_{1,D}$ of the quasi-dark state is instead approximately linear in the detuning and sits in between the JCM-like energies. An elementary perturbation theory calculation indeed yields $E_{n,D}\simeq\Delta\left(1 -{ \varepsilon ^2}/{4 g_\mathrm{eff}^2 n}\right)$, which agrees well with the exact value in the considered regime (relative error below $0.2\%$). Qualitatively similar results are obtained for higher excitation numbers $n\geq 2$, and in fact it is easy to see that the relative weight of $\varepsilon$ and $\Delta$ decreases with $n$.}\label{lowspectrum}
	\end{figure}

	We here present the analytical diagonalization of Hamiltonian \eqref{Hmany} for the case $N=2$ with equal couplings $g_1=g_2=g$, as well as the associated implications for the system's open dynamics. This instance of the model, illustrated in Fig.~\ref{Figlivelli}, is particularly transparent and fully captures the essence of our study. We verified that analytical solutions are also available for the cases $N=3,4$, using similar methods, but these are significantly more complex without adding fundamentally new insights. For $N=2$ the detunings are given by $\Delta_1=\Delta-\varepsilon/2$ and $\Delta_2=\Delta+\varepsilon/2$, respectively. Furthermore there are only one bright and one dark state given by $\ket{B}=\frac{\ket{e_1}+\ket{e_2}}{\sqrt{2}}$, $\ket{D}=\frac{\ket{e_1}-\ket{e_2}}{\sqrt2}$. In the radiation basis representation, the Hamiltonian may be recast as
	\begin{align}
	H&=H_0'+V',\label{Hprime}\\
	H_0'&=\omega_0\hat a^\dagger \hat a+\left(\omega_0+\Delta\right)(\proj{B}+\proj{D}),\label{H0prime}\\
	V'&=g_\mathrm{eff}\left(\hat a\,\kebra{B}{G}+\hat a^\dagger\kebra{G}{B}\right)-\frac{\varepsilon}{2}\left(\kebra{B}{D}+\kebra{D}{B}\right),\label{Vprime}
	\end{align}
	where in this case $g_\mathrm{eff}=\sqrt2 g$. As for the general case treated in the previous section, a JC interaction describes the coupling between the two levels $\{\ket{G},\ket{B}\}$ and the cavity field. However we can also see that the relative detuning of the sub-levels, $\varepsilon$, translates into a small coupling between $\ket{B}$ and $\ket{D}$, so that even the dark state $\ket{D}$ can indirectly influence the optical behaviour of the system. The ground state of $H$ is easily spotted as $\ket{G,0}$, with eingenvalue $E=0$. Exploting excitation number conservation, i.e. $[H,\hat{\mathcal{N}}_\mathrm{tot}]=0,$ where $\hat{\mathcal{N}}_\mathrm{tot}=\hat a^\dagger\hat a+\proj{B}+\proj{D}$, one can show that the remaining eigenvalues and eigenvectors of $H$ can be found by diagonalising the $3\times 3$ matrices ($n=1,2,\dots,\infty$)
	\begin{align}
	\mathbb{H}_n=\begin{pmatrix}
	n\omega_0 &  g_\mathrm{eff} \sqrt n& 0\\
	g_\mathrm{eff} \sqrt n& n\omega_0 +\Delta & -\varepsilon/2\\
	0&-\varepsilon/2 &n\omega_0+\Delta
	\end{pmatrix},\label{Hn}
	\end{align}
	which describe the Hamiltonian restricted to the subspace 
	\begin{equation}
	\left\{\hat{\mathcal{N}}_\mathrm{tot}=n\right\} \equiv \mathrm{Span}\Big\{\ket{G,n},\ket{B,n-1},\ket{D,n-1}\Big\}.\label{spanno}
	\end{equation}
	The behaviour of the eigenvalues of $\mathbb{H}_1$ is shown in Fig~\ref{lowspectrum}, as a function of the detuning $\Delta$. Even at the non-negligible ratio $\varepsilon/g_\mathrm{eff}=0.25$ used in the figure, the commonalities with a JCM spectrum are evident. Note also that the eigenvalues of $\mathbb{H}_n$ for $n\geq 2$ will follow the same qualitative structure, as a generic $\mathbb{H}_n$ may be obtained from $\mathbb{H}_1$ by a simple rescaling $g_\mathrm{eff}\to g_\mathrm{eff}\sqrt{n}$, $\omega_0\to n\omega_0$. While a general analytical diagonalization of Eq.~\eqref{Hn} is cumbersome, for the resonant case $\Delta=0$ one can obtain simple analytical expressions. For this case, the eigenvalues of $\mathbb{H}_n$ read
	\begin{align}
	E_{n,\pm}&=n\omega_0\pm g_\mathrm{eff}\sqrt{n+\frac{\varepsilon^2}{4g_\mathrm{eff}^2}},\\
	E_{n,D}&=n\omega_0.
	\end{align}
	The corresponding normalised eigenvectors $\ket{E_{n,\pm}},\ket{	E_{n,D}}$ are
	\begin{align}
	\ket{E_{n,\pm}}&=\frac{1}{\sqrt2}\left(\frac{2g_\mathrm{eff}\sqrt n}{\sqrt{4g_\mathrm{eff}^2n+\varepsilon^2}}\ket{G,n}\pm\ket{B,n-1} \right)\nonumber\\
	&-\frac{\varepsilon}{\sqrt{8g_\mathrm{eff}^2n+2\varepsilon^2}}\ket{D,n-1},\\
	\ket{E_{n,D}}&=\frac{\varepsilon}{\sqrt{4g_\mathrm{eff}^2n+\varepsilon^2}}\ket{G,n}+\frac{2g_\mathrm{eff}\sqrt n}{\sqrt{4g_\mathrm{eff}^2n+\varepsilon^2}}\ket{D,n-1}.
	\end{align}
	The above expressions suggest that in the regime $\varepsilon\ll g_\mathrm{eff}$ it is natural to identify $\{ \ket{E_{n,\pm}}, E_{n,\pm}\}$ as a weakly perturbed JC eigensystem (the latter is indeed recovered exactly for $\varepsilon\to0$). For brevity, we refer to those as the `JC model-type' eigenvalues and eigenvectors. The eigenstates $\ket{E_{n,D}}$, which instead feature the same energy levels of an empty cavity, may be understood as dressed quasi-dark states of the emitter, accompanied by approximately $n-1$ cavity photons. They indeed result from a hybridization of states $\ket{G,n}$ and $\ket{D,n-1}$, with the second state being dominant in the superposition. It is interesting to notice that the two states involved in such superposition are only coupled indirectly in the Hamiltonian, through the mediation of state $\ket{B,n-1}$.\\
	Exploiting the analytical results obtained here, it is possible to calculate the Hamiltonian dynamics of a generic initial state via standard methods. In fact, while our work focuses on the regime $\varepsilon\ll g_\mathrm{eff},$ we note that the above diagonalization procedure is valid for a general choice of the model parameters.
	
	\subsection{Qualitative analysis of the master equation for \texorpdfstring{$N=2$}{N=2}}\label{QualitativeN2}
	The closed expressions obtained so far are particularly helpful in analysing the open system behaviour of our multilevel model, as well as its relationship to the JC model. Since in this work we focus on the regime of low pumping (or low coherent driving) and negligible emitter decay, we can gain valuable intuition about the properties of the master equation \eqref{master} by setting $\gamma=\Lambda=\Omega=0$. First, we observe that the introduction of a photon loss rate $\kappa$ results in a finite lifetime for the excited states of $H$, such that eigenstates with $\hat{\mathcal{N}}_\mathrm{tot}=n$ will tend to decay towards states with $\hat{\mathcal{N}}_\mathrm{tot}=n-1$, until the ground state $\ket{G,0}$ is finally reached. For a given eigenstate $\ket{E_{j}}$, an estimate of the decay rate induced by photon loss can be obtained via the perturbation theory formula
	\begin{align}
	\Gamma_{j}\simeq\kappa\bra{E_j}\hat a^\dagger \hat a\ket{E_j},
	\end{align}
	which at second order in $\varepsilon$ yields ($n\geq1$)
	\begin{align}
	\Gamma_{nD}&\simeq\kappa\left(n-1+\frac{\varepsilon^2}{4ng_\mathrm{eff}^2}\right),\label{GammaD}\\
	\Gamma_{n\pm}&\simeq\kappa \left(n-\frac12-\frac{\varepsilon^2}{8ng_\mathrm{eff}^2}\right)\label{GammaPM}.
	\end{align}
	We may immediately notice that the lowest quasi-dark state, $\ket{E_{1D}}$, is long lived, while all the remaining excited states have approximately the same lifetime of either the JCM excited states or the Fock states of an empty and lossy cavity. Next, emitter dephasing will induce further broadening of the Hamiltonian energy levels, which can be estimated as
	\begin{align}
	\delta E_j\simeq\gamma_d\left(|\sprod{e_1}{E_j}|^2+|\sprod{e_2}{E_j}|^2\right).
	\end{align}
	Some straightforward algebra yields
	\begin{align}
	\delta E_{nD}&\simeq \gamma_d\left(1-\frac{\varepsilon^2}{4ng_\mathrm{eff}^2}\right),\\
	\delta E_{n\pm}&\simeq \frac{\gamma_d}{2}\left(1+\frac{\varepsilon^2}{4ng_\mathrm{eff}^2}\right),
	\end{align}
	where again we have expanded our results at second order in $\varepsilon$. In this case we can see that the quasi-dark states are the most affected, since they feature a larger overlap with the bare excited states of the emitter. Note also that the $\delta E_j$'s are not associated with a decay towards states of lower excitation number (more below). The exact broadening of the energy levels, induced by the combined effect of photon loss and emitter dephasing, can be obtained as $-2\mathrm{Im}(\tilde E_j)$, where $ \tilde E_j$ are the eigenvalues of the non-Hermitian operator 
	\begin{align}
	\tilde H=H-i\frac{\kappa}{2}\hat a^\dagger \hat a-\frac{\gamma_d}{2}\sum_k\proj{e_k }.
	\end{align}
	The resulting expressions are unwieldy, however we checked via a numerical optimization that $-2\mathrm{Im}(\tilde E_j)\simeq\Gamma_j+\delta E_j$, with a relative error below $\simeq3.1\%$ in the range $0\leq\varepsilon\leq0.25g_\mathrm{eff},0\leq\kappa\leq2g_\mathrm{eff},0\leq\gamma_d\leq 2g_\mathrm{eff}$, which is more than enough for our scopes.
	
	In order to gain qualitative understanding of photon emission processes in our system, we next look at the (squared) matrix elements of the annihilation operator $\hat a$, 
	\begin{align}
	A_{i\to f}=|\bra{E_f}\hat a\ket{E_i}|^2,
	\end{align}
	between an initial energy eigenstate $\ket{E_i}$ and a final one $\ket{E_f}$. The quantum jump approach to GKSL master equations \cite{Breuer} provides a useful interpretation for these quantities: $A_{i\to f}$ is proportional to the probability that the transition $\ket{E_i}\to\ket{E_f} $ occurs upon loss of a cavity photon [in the master equation \eqref{master}, such transitions are triggered by the term $\kappa\hat a\rho\hat a^\dagger$]. When the transition occurs, a photon is emitted into extra-cavity modes with (approximately) frequency $\omega_{fi}=E_f-E_i$ and bandwidth $\Gamma_i+\Gamma_f+\delta E_i+\delta E_f$.
	The $A_{i\to f}$ terms involving only JCM-type eigenstates do not present any particular surprises: they approximate the well-known JCM results for  $\varepsilon\ll g_\mathrm{eff}$, and converge to them for $\varepsilon\to0$. We thus omit such cases for brevity and focus only on the matrix elements involving quasi dark states, which yield the following expressions:
	\begin{align}
	A_{1D\to0}&=\frac{\varepsilon^2}{4g_\mathrm{eff}^2n+\varepsilon^2},\label{darkG}\\
	A_{nD\to n-1,D}&=n-\frac{4 g_\mathrm{eff}^2 n}{4 g_\mathrm{eff}^2 n+\varepsilon ^2},\label{darkdark}\\
	A_{nD\to n-1,\pm}&=0,\label{darkpm}\\
	A_{n\pm\to n-1,D}&=\frac{2 g_\mathrm{eff}^2 \varepsilon ^2}{\left(4 g_\mathrm{eff}^2 (n-1)+\varepsilon ^2\right) \left(4 g_\mathrm{eff}^2 n+\varepsilon ^2\right)}.\label{pmdark}
	\end{align}
	From Eqs.~\eqref{darkG}, \eqref{darkdark} and \eqref{darkpm} we notice that, once the system is in one of the quasi-dark states $\ket{E_{nD}},$ the loss of a cavity photon can only produce the quasi dark state $\ket{E_{n-1,D}}$ for $n\geq 2$, or $\ket{G,0}$ for $n=1$. This corresponds to the emission of a photon at the cavity frequency $\omega_0$, with the associated matrix element of the order $\sim n-1$. Assuming for the moment $\gamma_d=0$ (no dephasing), this implies that once the system has evolved into a quasi-dark state, it essentially reproduces the open dynamics of an empty cavity until state $\ket{1,D}$ is reached. From there the system will decay slowly towards the lowest state $\ket{G,0}$, resulting again in the emission of a photon at the cavity frequency $\omega_0$, but with a much narrower bandwidth $\Gamma_{1D}\simeq \kappa \varepsilon^2/4g_\mathrm{eff}^2$. On the other hand, from Eq.~\eqref{pmdark} we can also see a small probability that a JCM-like state decays into a quasi-dark state by loss of a cavity photon. This process results in the emission of photons at average frequency $\omega_0\pm g_\mathrm{eff}\sqrt{n}$: these exactly overlap with the JCM spectral resonances for $n=1$, while they provide novel emission peaks for $n\geq2$. Putting back finite dephasing $\gamma_d\neq0$ into the picture, we see that, in addition to introducing further broadenings of the levels, dephasing events are capable of interconverting states $\ket B$ and $\ket D$. More in detail, we find $|\bra{f}L\ket{i}|^2=\gamma_d/2$ for $L=\sqrt{\gamma_d}\proj{e_k}$ with $k=1,2$ and $i,f\in\{D,B\}$. Referring again to the quantum jump approach \cite{Breuer} this will swap the states $\ket B$ and $\ket D$ with probability $1/2$, if a dephasing event occurs.
	\\
	In summary, we have provided a detailed characterization of the processes occurring in our open system for the emblematic case $N=2,\ \Delta=\Lambda=\gamma=\Omega=0$. While the multilevel model can approximate all features of a JCM for $\varepsilon\ll g_\mathrm{eff}$, it displays additional phenomena due to the indirect interaction of dark states and cavity field. In the next section we explore these issues quantitatively via numerical simulations.

	\begin{figure*}[t!]
		\begin{center}
		\textbf{Single-excitation regime - $\kappa=g_\mathrm{eff}$}
		\includegraphics[trim=100 0 140 0,clip,width=\textwidth]{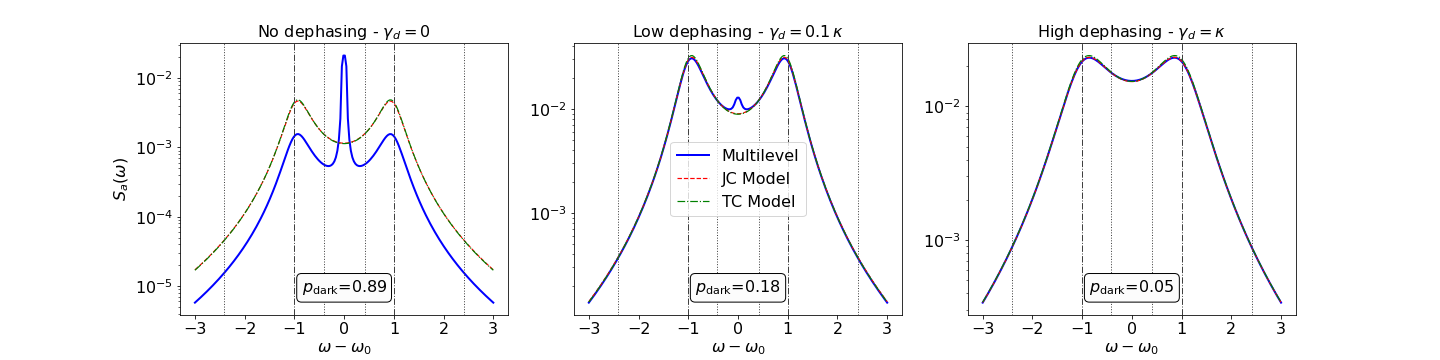}
		\textbf{~\\Bi-excitation regime - $\kappa=0.05g_\mathrm{eff}$}
		\includegraphics[trim=100 0 140 0,clip,width=\textwidth]{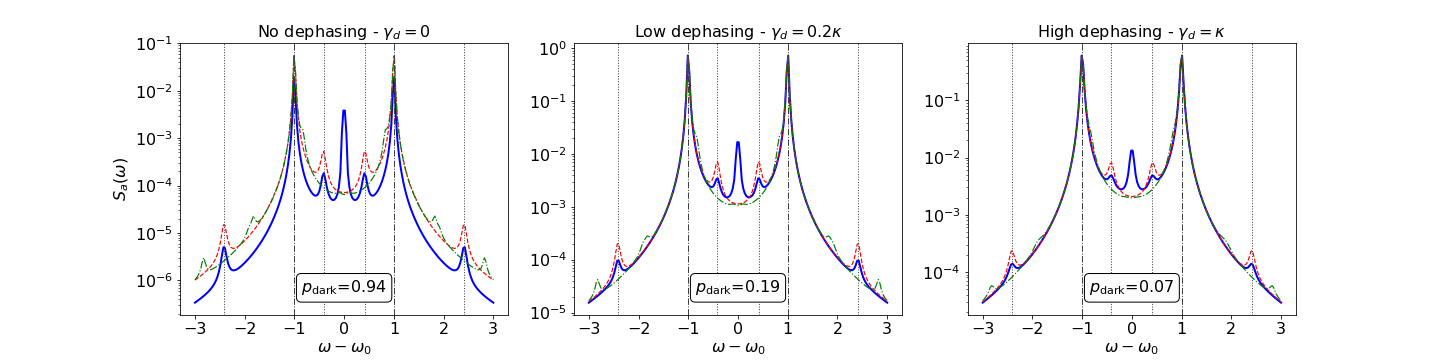}
		\caption{Comparison of cavity emission spectra, $S_a(\omega)$, as obtained for our multilevel model, versus reference JC and TC models. The sub-level energy spread is fixed at $\varepsilon=0.05g_\mathrm{eff}$ in all plots. Spectra are obtained under  incoherent emitter pumping, i.e. $\Lambda\neq0,\Omega=0$. The top plots in the ``single-excitation regime" are obtained for $\kappa=g_\mathrm{eff}$and  $\Lambda=0.01\kappa$, while the ``bi-excitation regime" at the bottom is for $\kappa =0.05g_\mathrm{eff}$ and $\Lambda=0.02\kappa$. Dephasing increases from left to right. Solid blue lines indicate our multi-level model, dashed red lines show the quantities of interest for a reference JC model, while green dot-dashed lines are for the TC model. The spectra of the TC and JC models are rescaled by a factor $(1-p_\mathrm{dark})$. Gray vertical lines indicate the resonances corresponding to the first (dot-dashed) and second rung (dotted) of the JC ladder.}\label{Snumerics}
		\end{center}
	\end{figure*}
	
	\section{Climbing The JC Ladder}\label{numericalsection}
	In the following numerical studies, we will investigate how closely the physics of our system \--- as relevant for typical quantum optics experiments \--- resembles a JCM with enhanced coupling constant $g_\mathrm{eff}\propto\sqrt{N}$. For the sake of comparison, we shall also display the behaviour of a TC model with $N$ atoms and the same effective coupling constant $g_\mathrm{eff}$ --- see Appendix~\ref{auxmodels} for further details. Simulations are done for $N=3$: this is one step up in complexity compared to the previous section, but simple enough that the resulting calculations can be handled by a standard computer.

	Our analysis focuses on two observables that are ubiquitous in both theoretical and experimental investigations of the JC model (proper definitions will be given below):
	\begin{itemize}
		\item The \textit{steady state spectrum} $S_a(\omega)$ of the cavity output field, when the system is excited via incoherent pumping of the emitter;
		\item The photon autocorrelation function $g^{(2)}(0)$, when the emitter is excited via a coherent driving field with variable detuning.
	\end{itemize}
	The first observable provides valuable information about the energy level structure of a quantum system, in particular highlighting its optically allowed transitions. Instead, the $g^{(2)}(0)$ may be seen as a witness of optical non-linearity: we will use it in our model to show that the $\sqrt N$ enhancement of the coupling constant, $g_{\mathrm{eff}}\propto\sqrt N$, does not come at the cost of ``linearising" the light-matter interaction (something that instead happens in the TC scenario as $N$ is increased, as we shall see below). To support this last point, we will indeed compare our autocorrelation functions with the $g^{(2)}(0)$ obtained for a reference ``coupled oscillator model" --- i.e. the quintessential linear system in our context (see Appendix~\ref{auxmodels} for details). Note that we are focusing only on  observables that can be measured via the cavity output field. 
	The latter has indeed the same definition in all models considered here, facilitating direct comparisons between them. On the other hand, emitter observables do not have an unambiguous correspondence in the different models, as they are defined on Hilbert spaces of different dimensions.\\
	In detail, according to standard input-output theory \cite{walls2007quantum}, the portion of the cavity output field that reaches our detector, say $\hat{E}_\mathrm{out}$ (scalar for simplicity) may be expressed as a function of the cavity annihilation operator $\hat a$ as follows:
	\begin{align}
	    \hat{E}_\mathrm{out}= E_\mathrm{c}\hat a +\hat{E}_\mathrm{vac},\label{Eout}
	\end{align}
	where $E_\mathrm{c}$ is a constant depending on the details of the cavity-plus-detector setup, while the operator $\hat{E}_\mathrm{vac}$ captures all ``vacuum noise" contributions from the detector's environment. Thanks to the assumption of a vacuum environment, which is justified for optical frequencies at room temperature, $\hat{E}_\mathrm{vac}$ does not contribute to any normally ordered observable constructed from $\hat{E}_\mathrm{out}$ --- and these are precisely the types of observables employed here. \\
	In the interest of brevity, we restrict ourselves to an experimentally desirable situation in which $\Delta=0$ and the hierarchy $\varepsilon \ll  \kappa\lesssim g_\mathrm{eff}$ holds. In other words we consider a strongly coupled system where the average transition frequency of the emitter is at resonance with the cavity, while the range of  sub-level detunings is within the bare cavity bandwidth. This scenario is indeed appropriate for the case of quantum dots embedded in plasmonic nanoresonators, as anticipated in Section~\ref{modelsection}. We shall also take the driving strengths ($\Omega$ for the coherent and  $\Lambda$ for the incoherent case) as the smallest parameters in the problem. This ensures that the driving will not induce significant modifications to the intrinsic resonances of the coupled system, nor to the conditions for achieving strong coupling \cite{ValleI,ValleII}. In this setting we focus on two representative parameter regimes: 
	\begin{itemize}
		\item[(i)] $g_\mathrm{eff}=\kappa$, or `single-excitation regime',
		\item[(ii)]$g_\mathrm{eff}=20\kappa$, or `bi-excitation regime'.
	\end{itemize}
	Loosely speaking, case (i) gives access to the `first rung' of the energy ladder, thus the resulting Physics should be similar to that of a pair of coupled oscillators in all models. Case (ii) instead allows us to explore the `second rung' of the ladder, where the anharmonic nature of the investigated models becomes more prominent. In both situations we shall analyse the role of emitter dephasing by considering three scenarios: no dephasing, low dephasing (i.e. one order of magnitude smaller than the cavity decay rate $\kappa)$ and high dephasing (i.e. dephasing and cavity decay rates of the same order).
	This is particularly relevant for quantum-dot implementations of our models, especially at room temperature, in which case emitter dephasing rates can be comparable to the cavity decay rate. Finally it is important note that we shall adopt a logarithmic scale in our plots, since the examined quantities display features of very different magnitudes. This implies that the weaker features displayed in our plots (e.g. the second-rung spectral resonances) may be challenging to detect in experiments.
	
	\subsection{Steady state spectra}
	As anticipated, our first set of numerical examples investigates the optical emission properties of our multilevel model under incoherent pumping of the emitter, as relevant in many quantum dot experiments also at ambient conditions. Specifically, we set $\Omega=0$ (no coherent driving) and $\Lambda\neq0$. In this setting we are interested in the steady state spectrum of the cavity output field, i.e.,
	
	\begin{align}
	S_a(\omega)&= 2\mathrm{Re}\left(\int_{0}^{\infty}\med{ \hat{E}_\mathrm{out}^\dagger(\tau)\hat{E}_\mathrm{out}(0)}_\infty e^{-i\omega\tau}{\rm d}\tau\right),\nonumber\\
	&\propto 2\mathrm{Re}\left(\int_{0}^{\infty}\med{\hat a^\dagger(\tau)\hat a(0)}_\infty e^{-i\omega\tau}{\rm d}\tau\right),
	\end{align}
	where Eq.~\eqref{Eout} has been used, the two-point correlation function $\med{\hat a^\dagger(\tau)\hat a(0)}_\infty$ is calculated via the quantum regression theorem \cite{Breuer}, and the expectation value is calculated on $\rho_\infty$, the steady state of the master equation. The latter is found by solving the linear system ${\cal L}\rho_\infty=0,\tr{\rho_\infty}=1,$ where the superoperator ${\cal L}$ is implicitly defined by rewriting the master equation \eqref{master}  as $\dot\rho={\cal L}\rho$. We perform such calculations numerically with Python, by truncating the Hilbert space of the cavity to a sufficiently high dimension to obtain numerical convergence.
	In addition to the cavity spectrum, we shall also be interested in the quantity
	\begin{align}
	p_\mathrm{dark}&=\sum_{k}\tr{\rho_\infty\proj{D_k}}\nonumber\\
	&=1-\tr{\rho_\infty(\proj{B}+\proj{G})},
	\end{align}
	i.e. the steady-state probability that the emitter has settled in a dark state (or a mixture thereof). The resulting spectra, in logarithmic scale, are plotted in Fig.~\ref{Snumerics} together with those of the reference TC and JC models (again, see Appendix~\ref{auxmodels} for details). For a fairer visual comparison between the three models we have also rescaled the spectrum of JC and TC models by a factor $1-p_\mathrm{dark}$. We indeed recall from Section~\ref{QualitativeN2} that the multilevel model cannot display JC physics once it gets ``stuck" into a dark state. \\
	
	In the single excitation regime, we find that all three models display the expected resonances at $\omega\simeq\omega_0\pm g_{\mathrm{eff}}$, but the multilevel model presents additional sharp features at $\omega\simeq\omega_0$. In the case of no dephasing, these resonances account for a significant portion of the total emitted energy, and correspondingly $p_\mathrm{dark}$ is close to one. We argue that the $\omega\simeq\omega_0$ features can be ascribed to the slow decay of the long-lived quasi-dark states of the single-excitation sector: referring to our discussion and notation in Section~\ref{QualitativeN2}, this involves transitions of the form $\ket{E_{1D}}\to\ket{G,0}$ (where $\ket{E_{1D}}$ can be any of the two quasi-dark states occurring for $N=3$). As dephasing is increased, going left to right in Fig.~\ref{Snumerics}, we find that the three models show increasing agreement, and at the same time $p_\mathrm{dark}$ is significantly reduced. Referring again to our analytical discussion in Section \ref{QualitativeN2}, we recall that dephasing is indeed able to swap bright and dark states. This opens a new decay channel for the quasi-dark states $\ket{E_{1D}}$: as dephasing is increased, we may reach a parameter regime where the quasi-dark states $\ket{E_{1D}}$ are more likely to decay by a process of the form $\ket{E_{1D}}\to\ket{E_{1\pm}}\to\ket{G,0}$, where the first step is due to dephasing, the second to photon emission. Hence, in such conditions JC-like features can become more prominent in the measured spectrum. At high dephasing ($\gamma_d=\kappa$), the dark-state features are effectively invisible in the emission spectrum. \\
	The bi-excitation scenario is where the multilevel model truly shines, compared to the reference TC model: we can see in the bottom of Fig.~\ref{Snumerics} that multilevel and JC models are indeed in agreement on the location of both single-excitation ($\omega\simeq\omega_0\pm g_\mathrm{eff}$, gray dot-dashed lines) and bi-excitation [$\omega\simeq\omega_0\pm (\sqrt2\pm1)g_\mathrm{eff}]$, gray dotted lines) resonances, while the bi-excitation peaks of the TC model are visibly shifted. Also in this case the multilevel model presents an additional resonance (or perhaps a collection of closely spaced resonances) at $\omega\simeq\omega_0.$ Referring again to the notation of Section~\ref{QualitativeN2}, these can be ascribed to two processes: the photons emitted in the $\ket{E_{1D}}\to\ket{G,0}$ transition, which is associated with a small decay rate, and those emitted in the ``second rung" process $\ket{E_{2D}}\to\ket{E_{1D}}$, which we recall is very similar to photon emission by an empty cavity, and hence has a singnificantly higher decay rate $\simeq \kappa$.
	We again notice that the $\omega\simeq\omega_0$ resonance can be weakened in the presence of dephasing, however it remains prominent even in the ``high dephasing" scenario. Our interpretation for this is the following: while dephasing of order $\kappa$ can suppress the $\ket{E_{1D}}\to\ket{G,0}$ process, as discussed above, the $\ket{E_{2D}}\to\ket{E_{1D}}$ decay channel remains a probable process since it is also associated with a decay rate of order $\kappa$. \\
	It is worth pointing out that if one goes beyond our symmetric pump hypothesis (populating all the emitter excited states with equal probability - see Section~\ref{MEQsec}) it may of course be possible to reduce the value of $p_\mathrm{dark}$ and hence the prominence of dark-state features in the cavity emission spectrum.
	
	To summarize, the results of this subsection tell us that the multilevel model is a viable system to test the characteristic $\propto \sqrt{n}g_\mathrm{eff}$ splitting of energy levels that is still the holy grail of experimental JC research. In this light, the main strength of the multilevel model is that the effective coupling $g_\mathrm{eff}$ may be significantly boosted by considering emitters with many closely spaced sublevels. 
	
	\subsection{Photon autocorrelation function}
		\begin{figure*}[t!]
		\begin{center}
		\textbf{Single-excitation regime - $\kappa=g$}
		\includegraphics[trim=100 0 140 0,clip,width=\textwidth]{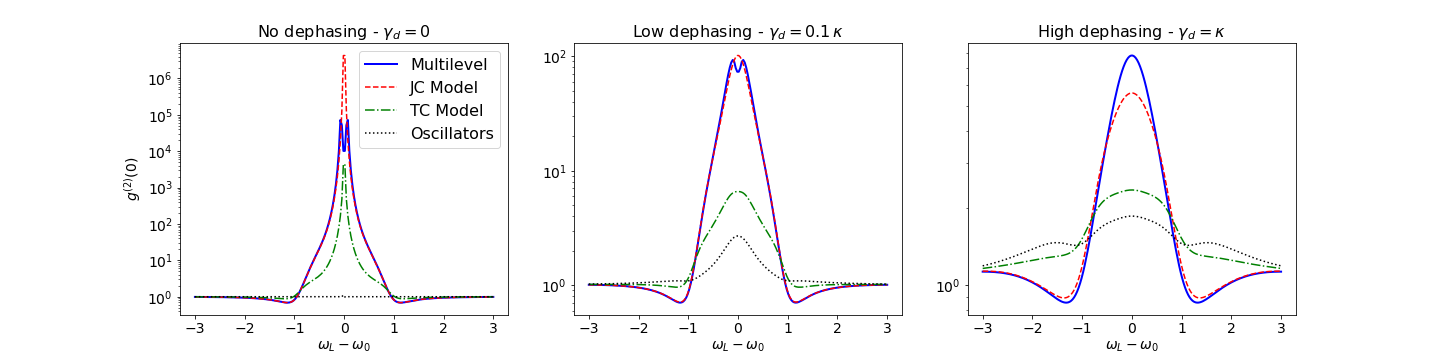}
		\textbf{~\\Bi-excitation regime - $\kappa=0.05g$}
		\includegraphics[trim=100 0 140 0,clip,width=\textwidth]{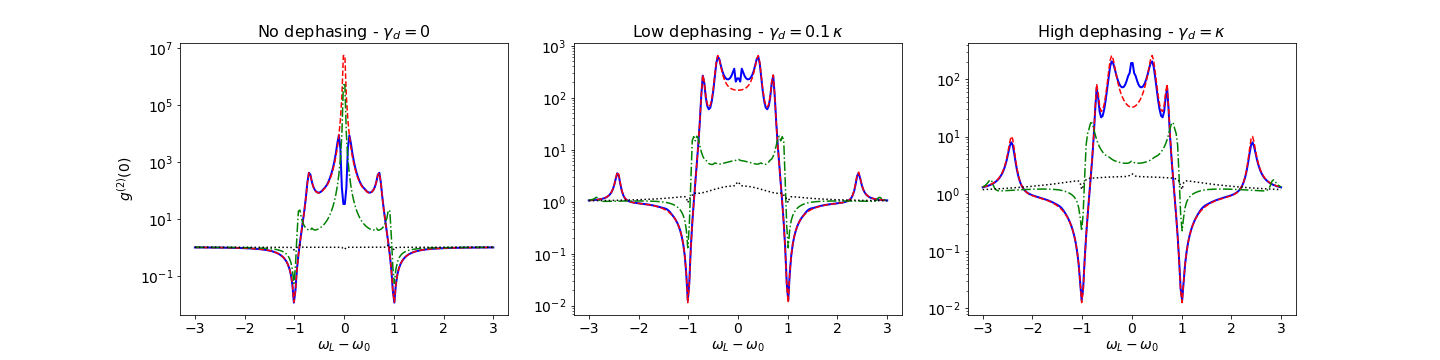}
		\caption{steady-state photon autocorrelation $g^{(2)}(0)$ for $N=3$ emitter sublevels and a sub-level energy spread fixed at $\varepsilon=0.05g_\mathrm{eff}$. We set the incoherent pump to zero and introduce a coherent cavity drive, corresponding to model parameters $\Lambda=0,\Omega=0.01g_\mathrm{eff}$. Dephasing increases from left to right as before. Solid blue lines = multi-level model; dashed red lines = reference JC model; dot-dashed green lines = reference TC model; dotted black lines = reference coupled oscillators model. For details on all the auxiliary models used here see Appendix~\ref{auxmodels}}\label{g2numerics}
	\end{center}
	\end{figure*}
	As a second step, we will look at the steady state behaviour of the zero-delay photon autocorrelation function, under weak coherent driving of the cavity. In all three models this quantity is defined as follows:
	\begin{align}
	g^{(2)}(0)&=\lim\limits_{t\to\infty}\frac{\av{\hat{E}_\mathrm{out}^\dagger \hat{E}_\mathrm{out}^\dagger \hat{E}_\mathrm{out}\hat{E}_\mathrm{out}}}{\av{\hat{E}_\mathrm{out}^\dagger \hat{E}_\mathrm{out}}^2}\nonumber\\
	&=\lim\limits_{t\to\infty}\frac{\av{\hat a^\dagger \hat a^\dagger \hat a\hat a}}{\av{\hat a^\dagger \hat a}^2},\label{g2def}
	\end{align}
	where again we made use of Eq.~\eqref{Eout} in order to obtain an expression that only features system operators. In detail, in this subsection we consider the master equation \eqref{master}, setting $\Lambda=0$, $\Omega\neq 0$, and vary the driving field frequency $\omega_L$. While the resulting master equation is explicitly time dependent, it is still possible to define a steady state by considering an interaction picture with respect to the Hamiltonian $H_L=\omega_L\hat a^\dagger \hat a+\omega_L\sum_{k=1}^n\proj{e_k}.$ This is a standard procedure in quantum optics and it amounts to replacing $V_\mathrm{drive}\to V_\mathrm{drive}^{(\mathrm{int})}=\Omega\left(\hat a +\hat a^\dagger \right)$ and 
	$H\to H^{(\mathrm{int})}=H-H_L$ in the master equation~\eqref{master}. In turn, this procedure implicitly defines a time independent superoperator $\mathcal {L}^{(\mathrm{int})}$ which does admit a steady state $\rho_\infty^{(\mathrm{int})}$. Such a steady state is again found numerically with the same process described in the previous section, and it can be easily shown that it provides the correct expectation values involved in Eq.~\eqref{g2def}. The obtained autocorrelation functions for multilevel, JC and TC models are displayed in Fig.~\ref{g2numerics} as functions of the driving field detuning. We note that in this case no rescaling of the TC and JC plots is required, since the $g^{2}(0)$ is normalised by definition.
	
	We observe that, excluding the region $\omega\simeq\omega_0$, the multilevel model does a good job at matching the JC model predictions, capturing both the position and magnitude of resonances and dips. These observations are true both in the single- and bi-excitation regimes. In contrast, the $g^{(2)}(0)$ of the TC model displays features that are significantly less pronounced than their JC counterparts (recall that we are using a logscale), as well as being shifted in frequency in the bi-excitation regime. This is compatible with the intuition that the TC model is ``more linear" than both the JC and multilevel ones. Finally, the coupled oscillator model is of course even farther off from the JC predictions. As expected, the latter indeed displays features that are several orders of magnitude smaller than those in all the other models, plus it satisfies $g^{(2)}(0)>1$ at all times.
	As anticipated, however, the multilevel model displays additional features in the region $\omega\simeq \omega_0$: at low or no dephasing a double-peak structure can be clearly seen, likely due to processes involving the quasi-dark states (we recall that for $N=3$ there are two quasi-dark states for each value of the excitation number $n\geq 1$). As dephasing is increased the prominence of these peaks is reduced, presumably due to the introduction of additional decay channels for the quasi-dark states (see also previous Subsection). However, dephasing also results in the appearance of a central peak in the $g^{(2)}(0)$, which may be due to an increased likelihood that a pair of driving photons excite a quasi-dark state of the system, and are later emitted at the frequency of an empty cavity by a process of the form $\ket{E_{2D}}\to\ket{E_{1D}}\to\ket{G,0}$. While a more rigorous analysis of these effects would be certainly worthwhile to confirm our intuitions, we feel that it goes beyond the scopes of this work.
	
	As anticipated, the $g^{(2)}(0)$ is one of the most commonly adopted witnesses of nonlinearity in emitter-cavity systems such as the ones we are investigating. The results of this subsection thus confirm that the multilevel model closely matches the nonlinear signatures of JC physics, and in this respect it clearly outperforms the reference TC model with the same values of $N$ and $g_\mathrm{eff}$.

	
	\section{Conclusions}\label{conclusions}
	We explored a novel route to realise the strong coupling regime of light-matter interaction, exploiting a multi-level emitter coupled to a single-mode electromagnetic field. When the $N$ excited sublevels of the emitter are nearly degenerate, we showed both analytically and numerically that 
	our proposed system can closely approximate many aspects of JC physics. The associated light-matter coupling constant scales as $g_\mathrm{eff}\propto\sqrt{N}$ and, crucially, the quantum nonlinearities of our model are not suppressed with increasing $N$. Indeed, we are able to observe the characteristic energy level splitting of order $\sim\sqrt{n}$, where $n$ is the number of combined (emitter + field) excitations, as well as the associated nonlinear signatures in emission spectra and photon autocorrelation functions. Moreover, we have been able to characterise the main effects associated with the additional levels of the emitter: the appearance of very interesting sharp resonances in the middle of the emission spectrum, associated with the slow decay of ``quasi dark states". While it may be possible to observe such resonances in an experiment, our analysis shows that these are quickly washed out in the presence of dephasing. Observing dark-state signatures would thus require  cryogenic temperatures.
	
	Our findings are particularly relevant in solid-state and plasmonic implementations of Cavity QED, where the emitters of choice are often multi-level quantum dots, and may pave the way towards the observation and exploitation of light-matter strong coupling at room temperature.
	\section{Acknowledgments}
	We thank M. G. Genoni for the useful discussion on modelling dephasing. Support by the Science Foundation Ireland (SFI) under grant 18/RP/6236 is gratefully acknowledged. TT acknowledges support from the University of Nottingham via a Nottingham Research Fellowship.  

\nocite{*}
\bibliography{lit}

\providecommand{\noopsort}[1]{}\providecommand{\singleletter}[1]{#1}%
\begin{thebibliography}{43}%
\makeatletter
\providecommand \@ifxundefined [1]{%
 \@ifx{#1\undefined}
}%
\providecommand \@ifnum [1]{%
 \ifnum #1\expandafter \@firstoftwo
 \else \expandafter \@secondoftwo
 \fi
}%
\providecommand \@ifx [1]{%
 \ifx #1\expandafter \@firstoftwo
 \else \expandafter \@secondoftwo
 \fi
}%
\providecommand \natexlab [1]{#1}%
\providecommand \enquote  [1]{``#1''}%
\providecommand \bibnamefont  [1]{#1}%
\providecommand \bibfnamefont [1]{#1}%
\providecommand \citenamefont [1]{#1}%
\providecommand \href@noop [0]{\@secondoftwo}%
\providecommand \href [0]{\begingroup \@sanitize@url \@href}%
\providecommand \@href[1]{\@@startlink{#1}\@@href}%
\providecommand \@@href[1]{\endgroup#1\@@endlink}%
\providecommand \@sanitize@url [0]{\catcode `\\12\catcode `\$12\catcode
  `\&12\catcode `\#12\catcode `\^12\catcode `\_12\catcode `\%12\relax}%
\providecommand \@@startlink[1]{}%
\providecommand \@@endlink[0]{}%
\providecommand \url  [0]{\begingroup\@sanitize@url \@url }%
\providecommand \@url [1]{\endgroup\@href {#1}{\urlprefix }}%
\providecommand \urlprefix  [0]{URL }%
\providecommand \Eprint [0]{\href }%
\providecommand \doibase [0]{https://doi.org/}%
\providecommand \selectlanguage [0]{\@gobble}%
\providecommand \bibinfo  [0]{\@secondoftwo}%
\providecommand \bibfield  [0]{\@secondoftwo}%
\providecommand \translation [1]{[#1]}%
\providecommand \BibitemOpen [0]{}%
\providecommand \bibitemStop [0]{}%
\providecommand \bibitemNoStop [0]{.\EOS\space}%
\providecommand \EOS [0]{\spacefactor3000\relax}%
\providecommand \BibitemShut  [1]{\csname bibitem#1\endcsname}%
\let\auto@bib@innerbib\@empty
\bibitem [{\citenamefont {Shore}\ and\ \citenamefont
  {Knight}(1993)}]{shore1993jaynes}%
  \BibitemOpen
  \bibfield  {author} {\bibinfo {author} {\bibfnamefont {B.~W.}\ \bibnamefont
  {Shore}}\ and\ \bibinfo {author} {\bibfnamefont {P.~L.}\ \bibnamefont
  {Knight}},\ }\bibfield  {title} {\bibinfo {title} {The jaynes-cummings
  model},\ }\href@noop {} {\bibfield  {journal} {\bibinfo  {journal} {Journal
  of Modern Optics}\ }\textbf {\bibinfo {volume} {40}},\ \bibinfo {pages}
  {1195} (\bibinfo {year} {1993})}\BibitemShut {NoStop}%
\bibitem [{\citenamefont {Haroche}\ and\ \citenamefont
  {Raimond}(2006)}]{haroche2006exploring}%
  \BibitemOpen
  \bibfield  {author} {\bibinfo {author} {\bibfnamefont {S.}~\bibnamefont
  {Haroche}}\ and\ \bibinfo {author} {\bibfnamefont {J.-M.}\ \bibnamefont
  {Raimond}},\ }\href@noop {} {\emph {\bibinfo {title} {Exploring the quantum:
  atoms, cavities, and photons}}}\ (\bibinfo  {publisher} {Oxford university
  press},\ \bibinfo {year} {2006})\BibitemShut {NoStop}%
\bibitem [{\citenamefont {Reiserer}\ \emph {et~al.}(2014)\citenamefont
  {Reiserer}, \citenamefont {Kalb}, \citenamefont {Rempe},\ and\ \citenamefont
  {Ritter}}]{reiserer_quantum_2014}%
  \BibitemOpen
  \bibfield  {author} {\bibinfo {author} {\bibfnamefont {A.}~\bibnamefont
  {Reiserer}}, \bibinfo {author} {\bibfnamefont {N.}~\bibnamefont {Kalb}},
  \bibinfo {author} {\bibfnamefont {G.}~\bibnamefont {Rempe}},\ and\ \bibinfo
  {author} {\bibfnamefont {S.}~\bibnamefont {Ritter}},\ }\bibfield  {title}
  {\bibinfo {title} {A quantum gate between a flying optical photon and a
  single trapped atom},\ }\href {https://doi.org/10.1038/nature13177}
  {\bibfield  {journal} {\bibinfo  {journal} {Nature}\ }\textbf {\bibinfo
  {volume} {508}},\ \bibinfo {pages} {237} (\bibinfo {year} {2014})},\ \bibinfo
  {note} {number: 7495 Publisher: Nature Publishing Group}\BibitemShut
  {NoStop}%
\bibitem [{\citenamefont {Biagioni}\ \emph {et~al.}(2012)\citenamefont
  {Biagioni}, \citenamefont {Huang},\ and\ \citenamefont {Hecht}}]{Biagioni}%
  \BibitemOpen
  \bibfield  {author} {\bibinfo {author} {\bibfnamefont {P.}~\bibnamefont
  {Biagioni}}, \bibinfo {author} {\bibfnamefont {J.-S.}\ \bibnamefont
  {Huang}},\ and\ \bibinfo {author} {\bibfnamefont {B.}~\bibnamefont {Hecht}},\
  }\bibfield  {title} {\bibinfo {title} {Nanoantennas for visible and infrared
  radiation},\ }\href {https://doi.org/10.1088/0034-4885/75/2/024402}
  {\bibfield  {journal} {\bibinfo  {journal} {Reports on Progress in Physics}\
  }\textbf {\bibinfo {volume} {75}},\ \bibinfo {pages} {024402} (\bibinfo
  {year} {2012})}\BibitemShut {NoStop}%
\bibitem [{\citenamefont {Walther}\ \emph {et~al.}(2006)\citenamefont
  {Walther}, \citenamefont {Varcoe}, \citenamefont {Englert},\ and\
  \citenamefont {Becker}}]{walther2006cavity}%
  \BibitemOpen
  \bibfield  {author} {\bibinfo {author} {\bibfnamefont {H.}~\bibnamefont
  {Walther}}, \bibinfo {author} {\bibfnamefont {B.~T.}\ \bibnamefont {Varcoe}},
  \bibinfo {author} {\bibfnamefont {B.-G.}\ \bibnamefont {Englert}},\ and\
  \bibinfo {author} {\bibfnamefont {T.}~\bibnamefont {Becker}},\ }\bibfield
  {title} {\bibinfo {title} {Cavity quantum electrodynamics},\ }\href@noop {}
  {\bibfield  {journal} {\bibinfo  {journal} {Reports on Progress in Physics}\
  }\textbf {\bibinfo {volume} {69}},\ \bibinfo {pages} {1325} (\bibinfo {year}
  {2006})}\BibitemShut {NoStop}%
\bibitem [{\citenamefont {Törmä}\ and\ \citenamefont
  {Barnes}(2015)}]{torma_strong_2015}%
  \BibitemOpen
  \bibfield  {author} {\bibinfo {author} {\bibfnamefont {P.}~\bibnamefont
  {Törmä}}\ and\ \bibinfo {author} {\bibfnamefont {W.~L.}\ \bibnamefont
  {Barnes}},\ }\bibfield  {title} {\bibinfo {title} {Strong coupling between
  surface plasmon polaritons and emitters: a review},\ }\href
  {https://doi.org/10.1088/0034-4885/78/1/013901} {\bibfield  {journal}
  {\bibinfo  {journal} {Reports on Progress in Physics}\ }\textbf {\bibinfo
  {volume} {78}},\ \bibinfo {pages} {013901} (\bibinfo {year}
  {2015})}\BibitemShut {NoStop}%
\bibitem [{\citenamefont {Dovzhenko}\ \emph {et~al.}(2018)\citenamefont
  {Dovzhenko}, \citenamefont {Ryabchuk}, \citenamefont {Rakovich},\ and\
  \citenamefont {Nabiev}}]{dovzhenko_lightmatter_2018}%
  \BibitemOpen
  \bibfield  {author} {\bibinfo {author} {\bibfnamefont {D.~S.}\ \bibnamefont
  {Dovzhenko}}, \bibinfo {author} {\bibfnamefont {S.~V.}\ \bibnamefont
  {Ryabchuk}}, \bibinfo {author} {\bibfnamefont {Y.~P.}\ \bibnamefont
  {Rakovich}},\ and\ \bibinfo {author} {\bibfnamefont {I.~R.}\ \bibnamefont
  {Nabiev}},\ }\bibfield  {title} {\bibinfo {title} {Light–matter interaction
  in the strong coupling regime: configurations, conditions, and
  applications},\ }\href {https://doi.org/10.1039/C7NR06917K} {\bibfield
  {journal} {\bibinfo  {journal} {Nanoscale}\ }\textbf {\bibinfo {volume}
  {10}},\ \bibinfo {pages} {3589} (\bibinfo {year} {2018})}\BibitemShut
  {NoStop}%
\bibitem [{\citenamefont {Yu}\ \emph {et~al.}(2019)\citenamefont {Yu},
  \citenamefont {Yuan}, \citenamefont {Xu}, \citenamefont {Yong}, \citenamefont
  {Qu},\ and\ \citenamefont {Song}}]{yu_strong_2019}%
  \BibitemOpen
  \bibfield  {author} {\bibinfo {author} {\bibfnamefont {X.}~\bibnamefont
  {Yu}}, \bibinfo {author} {\bibfnamefont {Y.}~\bibnamefont {Yuan}}, \bibinfo
  {author} {\bibfnamefont {J.}~\bibnamefont {Xu}}, \bibinfo {author}
  {\bibfnamefont {K.-T.}\ \bibnamefont {Yong}}, \bibinfo {author}
  {\bibfnamefont {J.}~\bibnamefont {Qu}},\ and\ \bibinfo {author}
  {\bibfnamefont {J.}~\bibnamefont {Song}},\ }\bibfield  {title} {\bibinfo
  {title} {Strong {Coupling} in {Microcavity} {Structures}: {Principle},
  {Design}, and {Practical} {Application}},\ }\href
  {https://doi.org/10.1002/lpor.201800219} {\bibfield  {journal} {\bibinfo
  {journal} {Laser \& Photonics Reviews}\ }\textbf {\bibinfo {volume} {13}},\
  \bibinfo {pages} {1800219} (\bibinfo {year} {2019})},\ \bibinfo {note} {zSCC:
  0000013}\BibitemShut {NoStop}%
\bibitem [{\citenamefont {Zengin}\ \emph {et~al.}(2013)\citenamefont {Zengin},
  \citenamefont {Johansson}, \citenamefont {Johansson}, \citenamefont
  {Antosiewicz}, \citenamefont {K\"{a}ll},\ and\ \citenamefont
  {Shegai}}]{g-aggregates}%
  \BibitemOpen
  \bibfield  {author} {\bibinfo {author} {\bibfnamefont {G.}~\bibnamefont
  {Zengin}}, \bibinfo {author} {\bibfnamefont {G.}~\bibnamefont {Johansson}},
  \bibinfo {author} {\bibfnamefont {P.}~\bibnamefont {Johansson}}, \bibinfo
  {author} {\bibfnamefont {T.~J.}\ \bibnamefont {Antosiewicz}}, \bibinfo
  {author} {\bibfnamefont {M.}~\bibnamefont {K\"{a}ll}},\ and\ \bibinfo
  {author} {\bibfnamefont {T.}~\bibnamefont {Shegai}},\ }\bibfield  {title}
  {\bibinfo {title} {Approaching the strong coupling limit in single plasmonic
  nanorods interacting with {J}-aggregates},\ }\href
  {https://doi.org/10.1038/srep03074} {\bibfield  {journal} {\bibinfo
  {journal} {Scientific Reports}\ }\textbf {\bibinfo {volume} {3}},\ \bibinfo
  {pages} {3074} (\bibinfo {year} {2013})},\ \bibinfo {note} {number: 1
  Publisher: Nature Publishing Group}\BibitemShut {NoStop}%
\bibitem [{\citenamefont {Tavis}\ and\ \citenamefont
  {Cummings}(1968)}]{tavis1968exact}%
  \BibitemOpen
  \bibfield  {author} {\bibinfo {author} {\bibfnamefont {M.}~\bibnamefont
  {Tavis}}\ and\ \bibinfo {author} {\bibfnamefont {F.~W.}\ \bibnamefont
  {Cummings}},\ }\bibfield  {title} {\bibinfo {title} {Exact solution for an
  n-molecule—radiation-field hamiltonian},\ }\href@noop {} {\bibfield
  {journal} {\bibinfo  {journal} {Physical Review}\ }\textbf {\bibinfo {volume}
  {170}},\ \bibinfo {pages} {379} (\bibinfo {year} {1968})}\BibitemShut
  {NoStop}%
\bibitem [{\citenamefont {Quesada}(2012)}]{Quesada-2012}%
  \BibitemOpen
  \bibfield  {author} {\bibinfo {author} {\bibfnamefont {N.}~\bibnamefont
  {Quesada}},\ }\bibfield  {title} {\bibinfo {title} {Strong coupling of two
  quantum emitters to a single light mode: {The} dissipative {Tavis}-{Cummings}
  ladder},\ }\href {https://doi.org/10.1103/PhysRevA.86.013836} {\bibfield
  {journal} {\bibinfo  {journal} {Physical Review A}\ }\textbf {\bibinfo
  {volume} {86}},\ \bibinfo {pages} {013836} (\bibinfo {year} {2012})},\
  \bibinfo {note} {publisher: American Physical Society}\BibitemShut {NoStop}%
\bibitem [{\citenamefont {Dicke}(1954)}]{DickeOriginal}%
  \BibitemOpen
  \bibfield  {author} {\bibinfo {author} {\bibfnamefont {R.~H.}\ \bibnamefont
  {Dicke}},\ }\bibfield  {title} {\bibinfo {title} {Coherence in spontaneous
  radiation processes},\ }\href {https://doi.org/10.1103/PhysRev.93.99}
  {\bibfield  {journal} {\bibinfo  {journal} {Phys. Rev.}\ }\textbf {\bibinfo
  {volume} {93}},\ \bibinfo {pages} {99} (\bibinfo {year} {1954})}\BibitemShut
  {NoStop}%
\bibitem [{\citenamefont {Wers\"{a}ll}\ \emph {et~al.}(2017)\citenamefont
  {Wers\"{a}ll}, \citenamefont {Cuadra}, \citenamefont {Antosiewicz},
  \citenamefont {Balci},\ and\ \citenamefont {Shegai}}]{molecular}%
  \BibitemOpen
  \bibfield  {author} {\bibinfo {author} {\bibfnamefont {M.}~\bibnamefont
  {Wers\"{a}ll}}, \bibinfo {author} {\bibfnamefont {J.}~\bibnamefont {Cuadra}},
  \bibinfo {author} {\bibfnamefont {T.~J.}\ \bibnamefont {Antosiewicz}},
  \bibinfo {author} {\bibfnamefont {S.}~\bibnamefont {Balci}},\ and\ \bibinfo
  {author} {\bibfnamefont {T.}~\bibnamefont {Shegai}},\ }\bibfield  {title}
  {\bibinfo {title} {Observation of {Mode} {Splitting} in {Photoluminescence}
  of {Individual} {Plasmonic} {Nanoparticles} {Strongly} {Coupled} to
  {Molecular} {Excitons}},\ }\href
  {https://doi.org/10.1021/acs.nanolett.6b04659} {\bibfield  {journal}
  {\bibinfo  {journal} {Nano Letters}\ }\textbf {\bibinfo {volume} {17}},\
  \bibinfo {pages} {551} (\bibinfo {year} {2017})},\ \bibinfo {note}
  {publisher: American Chemical Society}\BibitemShut {NoStop}%
\bibitem [{\citenamefont {Baranov}\ \emph {et~al.}(2020)\citenamefont
  {Baranov}, \citenamefont {Munkhbat}, \citenamefont {Zhukova}, \citenamefont
  {Bisht}, \citenamefont {Canales}, \citenamefont {Rousseaux}, \citenamefont
  {Johansson}, \citenamefont {Antosiewicz},\ and\ \citenamefont
  {Shegai}}]{baranov_ultrastrong_2020}%
  \BibitemOpen
  \bibfield  {author} {\bibinfo {author} {\bibfnamefont {D.~G.}\ \bibnamefont
  {Baranov}}, \bibinfo {author} {\bibfnamefont {B.}~\bibnamefont {Munkhbat}},
  \bibinfo {author} {\bibfnamefont {E.}~\bibnamefont {Zhukova}}, \bibinfo
  {author} {\bibfnamefont {A.}~\bibnamefont {Bisht}}, \bibinfo {author}
  {\bibfnamefont {A.}~\bibnamefont {Canales}}, \bibinfo {author} {\bibfnamefont
  {B.}~\bibnamefont {Rousseaux}}, \bibinfo {author} {\bibfnamefont
  {G.}~\bibnamefont {Johansson}}, \bibinfo {author} {\bibfnamefont {T.~J.}\
  \bibnamefont {Antosiewicz}},\ and\ \bibinfo {author} {\bibfnamefont
  {T.}~\bibnamefont {Shegai}},\ }\bibfield  {title} {\bibinfo {title}
  {Ultrastrong coupling between nanoparticle plasmons and cavity photons at
  ambient conditions},\ }\href {https://doi.org/10.1038/s41467-020-16524-x}
  {\bibfield  {journal} {\bibinfo  {journal} {Nature Communications}\ }\textbf
  {\bibinfo {volume} {11}},\ \bibinfo {pages} {2715} (\bibinfo {year}
  {2020})}\BibitemShut {NoStop}%
\bibitem [{\citenamefont {Geisler}\ \emph {et~al.}(2019)\citenamefont
  {Geisler}, \citenamefont {Cui}, \citenamefont {Wang}, \citenamefont
  {Rindzevicius}, \citenamefont {Gammelgaard}, \citenamefont {Jessen},
  \citenamefont {Gon\c{c}alves}, \citenamefont {Todisco}, \citenamefont
  {B\o~ggild}, \citenamefont {Boisen}, \citenamefont {Wubs}, \citenamefont
  {Mortensen}, \citenamefont {Xiao},\ and\ \citenamefont
  {Stenger}}]{Geisler-2019}%
  \BibitemOpen
  \bibfield  {author} {\bibinfo {author} {\bibfnamefont {M.}~\bibnamefont
  {Geisler}}, \bibinfo {author} {\bibfnamefont {X.}~\bibnamefont {Cui}},
  \bibinfo {author} {\bibfnamefont {J.}~\bibnamefont {Wang}}, \bibinfo {author}
  {\bibfnamefont {T.}~\bibnamefont {Rindzevicius}}, \bibinfo {author}
  {\bibfnamefont {L.}~\bibnamefont {Gammelgaard}}, \bibinfo {author}
  {\bibfnamefont {B.~S.}\ \bibnamefont {Jessen}}, \bibinfo {author}
  {\bibfnamefont {P.~A.~D.}\ \bibnamefont {Gon\c{c}alves}}, \bibinfo {author}
  {\bibfnamefont {F.}~\bibnamefont {Todisco}}, \bibinfo {author} {\bibfnamefont
  {P.}~\bibnamefont {B\o~ggild}}, \bibinfo {author} {\bibfnamefont
  {A.}~\bibnamefont {Boisen}}, \bibinfo {author} {\bibfnamefont
  {M.}~\bibnamefont {Wubs}}, \bibinfo {author} {\bibfnamefont {N.~A.}\
  \bibnamefont {Mortensen}}, \bibinfo {author} {\bibfnamefont {S.}~\bibnamefont
  {Xiao}},\ and\ \bibinfo {author} {\bibfnamefont {N.}~\bibnamefont
  {Stenger}},\ }\bibfield  {title} {\bibinfo {title} {Single-{Crystalline}
  {Gold} {Nanodisks} on {WS2} {Mono}- and {Multilayers} for {Strong} {Coupling}
  at {Room} {Temperature}},\ }\href
  {https://doi.org/10.1021/acsphotonics.8b01766} {\bibfield  {journal}
  {\bibinfo  {journal} {ACS Photonics}\ }\textbf {\bibinfo {volume} {6}},\
  \bibinfo {pages} {994} (\bibinfo {year} {2019})},\ \bibinfo {note}
  {publisher: American Chemical Society}\BibitemShut {NoStop}%
\bibitem [{\citenamefont {Moilanen}\ \emph {et~al.}(2018)\citenamefont
  {Moilanen}, \citenamefont {Hakala},\ and\ \citenamefont
  {Törmä}}]{moilanen_active_2018}%
  \BibitemOpen
  \bibfield  {author} {\bibinfo {author} {\bibfnamefont {A.~J.}\ \bibnamefont
  {Moilanen}}, \bibinfo {author} {\bibfnamefont {T.~K.}\ \bibnamefont
  {Hakala}},\ and\ \bibinfo {author} {\bibfnamefont {P.}~\bibnamefont
  {Törmä}},\ }\bibfield  {title} {\bibinfo {title} {Active {Control} of
  {Surface} {Plasmon}–{Emitter} {Strong} {Coupling}},\ }\href
  {https://doi.org/10.1021/acsphotonics.7b00655} {\bibfield  {journal}
  {\bibinfo  {journal} {ACS Photonics}\ }\textbf {\bibinfo {volume} {5}},\
  \bibinfo {pages} {54} (\bibinfo {year} {2018})}\BibitemShut {NoStop}%
\bibitem [{\citenamefont {Santhosh}\ \emph {et~al.}(2016)\citenamefont
  {Santhosh}, \citenamefont {Bitton}, \citenamefont {Chuntonov},\ and\
  \citenamefont {Haran}}]{Santhosh}%
  \BibitemOpen
  \bibfield  {author} {\bibinfo {author} {\bibfnamefont {K.}~\bibnamefont
  {Santhosh}}, \bibinfo {author} {\bibfnamefont {O.}~\bibnamefont {Bitton}},
  \bibinfo {author} {\bibfnamefont {L.}~\bibnamefont {Chuntonov}},\ and\
  \bibinfo {author} {\bibfnamefont {G.}~\bibnamefont {Haran}},\ }\bibfield
  {title} {\bibinfo {title} {Vacuum {Rabi} splitting in a plasmonic cavity at
  the single quantum emitter limit},\ }\href
  {https://doi.org/10.1038/ncomms11823} {\bibfield  {journal} {\bibinfo
  {journal} {Nature Communications}\ }\textbf {\bibinfo {volume} {7}},\
  \bibinfo {pages} {ncomms11823} (\bibinfo {year} {2016})},\ \bibinfo {note}
  {number: 1 Publisher: Nature Publishing Group}\BibitemShut {NoStop}%
\bibitem [{\citenamefont {Chikkaraddy}\ \emph {et~al.}(2016)\citenamefont
  {Chikkaraddy}, \citenamefont {de~Nijs}, \citenamefont {Benz}, \citenamefont
  {Barrow}, \citenamefont {Scherman}, \citenamefont {Rosta}, \citenamefont
  {Demetriadou}, \citenamefont {Fox}, \citenamefont {Hess},\ and\ \citenamefont
  {Baumberg}}]{Baumberg}%
  \BibitemOpen
  \bibfield  {author} {\bibinfo {author} {\bibfnamefont {R.}~\bibnamefont
  {Chikkaraddy}}, \bibinfo {author} {\bibfnamefont {B.}~\bibnamefont
  {de~Nijs}}, \bibinfo {author} {\bibfnamefont {F.}~\bibnamefont {Benz}},
  \bibinfo {author} {\bibfnamefont {S.~J.}\ \bibnamefont {Barrow}}, \bibinfo
  {author} {\bibfnamefont {O.~A.}\ \bibnamefont {Scherman}}, \bibinfo {author}
  {\bibfnamefont {E.}~\bibnamefont {Rosta}}, \bibinfo {author} {\bibfnamefont
  {A.}~\bibnamefont {Demetriadou}}, \bibinfo {author} {\bibfnamefont
  {P.}~\bibnamefont {Fox}}, \bibinfo {author} {\bibfnamefont {O.}~\bibnamefont
  {Hess}},\ and\ \bibinfo {author} {\bibfnamefont {J.~J.}\ \bibnamefont
  {Baumberg}},\ }\bibfield  {title} {\bibinfo {title} {Single-molecule strong
  coupling at room temperature in plasmonic nanocavities},\ }\href
  {https://doi.org/10.1038/nature17974} {\bibfield  {journal} {\bibinfo
  {journal} {Nature}\ }\textbf {\bibinfo {volume} {535}},\ \bibinfo {pages}
  {127} (\bibinfo {year} {2016})},\ \bibinfo {note} {number: 7610 Publisher:
  Nature Publishing Group}\BibitemShut {NoStop}%
\bibitem [{\citenamefont {Gro\ss}\ \emph {et~al.}(2018)\citenamefont {Gro\ss},
  \citenamefont {Hamm}, \citenamefont {Tufarelli}, \citenamefont {Hess},\ and\
  \citenamefont {Hecht}}]{our_paper}%
  \BibitemOpen
  \bibfield  {author} {\bibinfo {author} {\bibfnamefont {H.}~\bibnamefont
  {Gro\ss}}, \bibinfo {author} {\bibfnamefont {J.~M.}\ \bibnamefont {Hamm}},
  \bibinfo {author} {\bibfnamefont {T.}~\bibnamefont {Tufarelli}}, \bibinfo
  {author} {\bibfnamefont {O.}~\bibnamefont {Hess}},\ and\ \bibinfo {author}
  {\bibfnamefont {B.}~\bibnamefont {Hecht}},\ }\bibfield  {title} {\bibinfo
  {title} {Near-field strong coupling of single quantum dots},\ }\href
  {https://doi.org/10.1126/sciadv.aar4906} {\bibfield  {journal} {\bibinfo
  {journal} {Science Advances}\ }\textbf {\bibinfo {volume} {4}},\ \bibinfo
  {pages} {eaar4906} (\bibinfo {year} {2018})}\BibitemShut {NoStop}%
\bibitem [{\citenamefont {Leng}\ \emph {et~al.}(2018)\citenamefont {Leng},
  \citenamefont {Szychowski}, \citenamefont {Daniel},\ and\ \citenamefont
  {Pelton}}]{leng_strong_2018}%
  \BibitemOpen
  \bibfield  {author} {\bibinfo {author} {\bibfnamefont {H.}~\bibnamefont
  {Leng}}, \bibinfo {author} {\bibfnamefont {B.}~\bibnamefont {Szychowski}},
  \bibinfo {author} {\bibfnamefont {M.-C.}\ \bibnamefont {Daniel}},\ and\
  \bibinfo {author} {\bibfnamefont {M.}~\bibnamefont {Pelton}},\ }\bibfield
  {title} {\bibinfo {title} {Strong coupling and induced transparency at room
  temperature with single quantum dots and gap plasmons},\ }\bibfield
  {journal} {\bibinfo  {journal} {Nature Communications}\ }\textbf {\bibinfo
  {volume} {9}},\ \href {https://doi.org/10.1038/s41467-018-06450-4}
  {10.1038/s41467-018-06450-4} (\bibinfo {year} {2018})\BibitemShut {NoStop}%
\bibitem [{\citenamefont {Park}\ \emph {et~al.}(2019)\citenamefont {Park},
  \citenamefont {May}, \citenamefont {Leng}, \citenamefont {Wang},
  \citenamefont {Kropp}, \citenamefont {Gougousi}, \citenamefont {Pelton},\
  and\ \citenamefont {Raschke}}]{park_tip-enhanced_2019}%
  \BibitemOpen
  \bibfield  {author} {\bibinfo {author} {\bibfnamefont {K.-D.}\ \bibnamefont
  {Park}}, \bibinfo {author} {\bibfnamefont {M.~A.}\ \bibnamefont {May}},
  \bibinfo {author} {\bibfnamefont {H.}~\bibnamefont {Leng}}, \bibinfo {author}
  {\bibfnamefont {J.}~\bibnamefont {Wang}}, \bibinfo {author} {\bibfnamefont
  {J.~A.}\ \bibnamefont {Kropp}}, \bibinfo {author} {\bibfnamefont
  {T.}~\bibnamefont {Gougousi}}, \bibinfo {author} {\bibfnamefont
  {M.}~\bibnamefont {Pelton}},\ and\ \bibinfo {author} {\bibfnamefont {M.~B.}\
  \bibnamefont {Raschke}},\ }\bibfield  {title} {\bibinfo {title} {Tip-enhanced
  strong coupling spectroscopy, imaging, and control of a single quantum
  emitter},\ }\href {https://doi.org/10.1126/sciadv.aav5931} {\bibfield
  {journal} {\bibinfo  {journal} {Science Advances}\ }\textbf {\bibinfo
  {volume} {5}},\ \bibinfo {pages} {eaav5931} (\bibinfo {year}
  {2019})}\BibitemShut {NoStop}%
\bibitem [{\citenamefont {Kasprzak}\ \emph {et~al.}(2010)\citenamefont
  {Kasprzak}, \citenamefont {Reitzenstein}, \citenamefont {Muljarov},
  \citenamefont {Kistner}, \citenamefont {Schneider}, \citenamefont {Strauss},
  \citenamefont {H\"{o}fling}, \citenamefont {Forchel},\ and\ \citenamefont
  {Langbein}}]{Kasprzak-2010}%
  \BibitemOpen
  \bibfield  {author} {\bibinfo {author} {\bibfnamefont {J.}~\bibnamefont
  {Kasprzak}}, \bibinfo {author} {\bibfnamefont {S.}~\bibnamefont
  {Reitzenstein}}, \bibinfo {author} {\bibfnamefont {E.~A.}\ \bibnamefont
  {Muljarov}}, \bibinfo {author} {\bibfnamefont {C.}~\bibnamefont {Kistner}},
  \bibinfo {author} {\bibfnamefont {C.}~\bibnamefont {Schneider}}, \bibinfo
  {author} {\bibfnamefont {M.}~\bibnamefont {Strauss}}, \bibinfo {author}
  {\bibfnamefont {S.}~\bibnamefont {H\"{o}fling}}, \bibinfo {author}
  {\bibfnamefont {A.}~\bibnamefont {Forchel}},\ and\ \bibinfo {author}
  {\bibfnamefont {W.}~\bibnamefont {Langbein}},\ }\bibfield  {title} {\bibinfo
  {title} {Up on the {Jaynes}-{Cummings} ladder of a quantum-dot/microcavity
  system},\ }\href {https://doi.org/10.1038/nmat2717} {\bibfield  {journal}
  {\bibinfo  {journal} {Nature Materials}\ }\textbf {\bibinfo {volume} {9}},\
  \bibinfo {pages} {304} (\bibinfo {year} {2010})},\ \bibinfo {note} {number: 4
  Publisher: Nature Publishing Group}\BibitemShut {NoStop}%
\bibitem [{\citenamefont {Fink}\ \emph {et~al.}(2008)\citenamefont {Fink},
  \citenamefont {G\"{o}ppl}, \citenamefont {Baur}, \citenamefont {Bianchetti},
  \citenamefont {Leek}, \citenamefont {Blais},\ and\ \citenamefont
  {Wallraff}}]{Fink-2008}%
  \BibitemOpen
  \bibfield  {author} {\bibinfo {author} {\bibfnamefont {J.~M.}\ \bibnamefont
  {Fink}}, \bibinfo {author} {\bibfnamefont {M.}~\bibnamefont {G\"{o}ppl}},
  \bibinfo {author} {\bibfnamefont {M.}~\bibnamefont {Baur}}, \bibinfo {author}
  {\bibfnamefont {R.}~\bibnamefont {Bianchetti}}, \bibinfo {author}
  {\bibfnamefont {P.~J.}\ \bibnamefont {Leek}}, \bibinfo {author}
  {\bibfnamefont {A.}~\bibnamefont {Blais}},\ and\ \bibinfo {author}
  {\bibfnamefont {A.}~\bibnamefont {Wallraff}},\ }\bibfield  {title} {\bibinfo
  {title} {Climbing the {Jaynes}–{Cummings} ladder and observing its
  nonlinearity in a cavity {QED} system},\ }\href
  {https://doi.org/10.1038/nature07112} {\bibfield  {journal} {\bibinfo
  {journal} {Nature}\ }\textbf {\bibinfo {volume} {454}},\ \bibinfo {pages}
  {315} (\bibinfo {year} {2008})},\ \bibinfo {note} {number: 7202 Publisher:
  Nature Publishing Group}\BibitemShut {NoStop}%
\bibitem [{\citenamefont {Reithmaier}\ \emph {et~al.}(2004)\citenamefont
  {Reithmaier}, \citenamefont {S\c{e}k}, \citenamefont {L\"{o}ffler},
  \citenamefont {Hofmann}, \citenamefont {Kuhn}, \citenamefont {Reitzenstein},
  \citenamefont {Keldysh}, \citenamefont {Kulakovskii}, \citenamefont
  {Reinecke},\ and\ \citenamefont {Forchel}}]{Reithmaier}%
  \BibitemOpen
  \bibfield  {author} {\bibinfo {author} {\bibfnamefont {J.~P.}\ \bibnamefont
  {Reithmaier}}, \bibinfo {author} {\bibfnamefont {G.}~\bibnamefont {S\c{e}k}},
  \bibinfo {author} {\bibfnamefont {A.}~\bibnamefont {L\"{o}ffler}}, \bibinfo
  {author} {\bibfnamefont {C.}~\bibnamefont {Hofmann}}, \bibinfo {author}
  {\bibfnamefont {S.}~\bibnamefont {Kuhn}}, \bibinfo {author} {\bibfnamefont
  {S.}~\bibnamefont {Reitzenstein}}, \bibinfo {author} {\bibfnamefont {L.~V.}\
  \bibnamefont {Keldysh}}, \bibinfo {author} {\bibfnamefont {V.~D.}\
  \bibnamefont {Kulakovskii}}, \bibinfo {author} {\bibfnamefont {T.~L.}\
  \bibnamefont {Reinecke}},\ and\ \bibinfo {author} {\bibfnamefont
  {A.}~\bibnamefont {Forchel}},\ }\bibfield  {title} {\bibinfo {title} {Strong
  coupling in a single quantum dot-semiconductor microcavity system},\ }\href
  {https://doi.org/10.1038/nature02969} {\bibfield  {journal} {\bibinfo
  {journal} {Nature}\ }\textbf {\bibinfo {volume} {432}},\ \bibinfo {pages}
  {197} (\bibinfo {year} {2004})},\ \bibinfo {note} {number: 7014 Publisher:
  Nature Publishing Group}\BibitemShut {NoStop}%
\bibitem [{\citenamefont {Madsen}\ \emph {et~al.}(2016)\citenamefont {Madsen},
  \citenamefont {Lehmann},\ and\ \citenamefont {Lodahl}}]{Madsen}%
  \BibitemOpen
  \bibfield  {author} {\bibinfo {author} {\bibfnamefont {K.~H.}\ \bibnamefont
  {Madsen}}, \bibinfo {author} {\bibfnamefont {T.~B.}\ \bibnamefont
  {Lehmann}},\ and\ \bibinfo {author} {\bibfnamefont {P.}~\bibnamefont
  {Lodahl}},\ }\bibfield  {title} {\bibinfo {title} {Role of multilevel states
  on quantum-dot emission in photonic-crystal cavities},\ }\href
  {https://doi.org/10.1103/PhysRevB.94.235301} {\bibfield  {journal} {\bibinfo
  {journal} {Physical Review B}\ }\textbf {\bibinfo {volume} {94}},\ \bibinfo
  {pages} {235301} (\bibinfo {year} {2016})},\ \bibinfo {note} {publisher:
  American Physical Society}\BibitemShut {NoStop}%
\bibitem [{\citenamefont {Liu}\ and\ \citenamefont
  {Guyot-Sionnest}(2010)}]{Liu-2010}%
  \BibitemOpen
  \bibfield  {author} {\bibinfo {author} {\bibfnamefont {H.}~\bibnamefont
  {Liu}}\ and\ \bibinfo {author} {\bibfnamefont {P.}~\bibnamefont
  {Guyot-Sionnest}},\ }\bibfield  {title} {\bibinfo {title} {Photoluminescence
  {Lifetime} of {Lead} {Selenide} {Colloidal} {Quantum} {Dots}},\ }\href
  {https://doi.org/10.1021/jp105818e} {\bibfield  {journal} {\bibinfo
  {journal} {The Journal of Physical Chemistry C}\ }\textbf {\bibinfo {volume}
  {114}},\ \bibinfo {pages} {14860} (\bibinfo {year} {2010})},\ \bibinfo {note}
  {publisher: American Chemical Society}\BibitemShut {NoStop}%
\bibitem [{\citenamefont {Hens}\ and\ \citenamefont
  {Moreels}(2012)}]{Hens-2012}%
  \BibitemOpen
  \bibfield  {author} {\bibinfo {author} {\bibfnamefont {Z.}~\bibnamefont
  {Hens}}\ and\ \bibinfo {author} {\bibfnamefont {I.}~\bibnamefont {Moreels}},\
  }\bibfield  {title} {\bibinfo {title} {Light absorption by colloidal
  semiconductor quantum dots},\ }\href {https://doi.org/10.1039/C2JM30760J}
  {\bibfield  {journal} {\bibinfo  {journal} {Journal of Materials Chemistry}\
  }\textbf {\bibinfo {volume} {22}},\ \bibinfo {pages} {10406} (\bibinfo {year}
  {2012})},\ \bibinfo {note} {publisher: The Royal Society of
  Chemistry}\BibitemShut {NoStop}%
\bibitem [{\citenamefont {Tserkezis}\ \emph {et~al.}(2020)\citenamefont
  {Tserkezis}, \citenamefont {Fern{\'{a}}ndez-Dom{\'{\i}}nguez}, \citenamefont
  {Gon{\c{c}}alves}, \citenamefont {Todisco}, \citenamefont {Cox},
  \citenamefont {Busch}, \citenamefont {Stenger}, \citenamefont {Bozhevolnyi},
  \citenamefont {Mortensen},\ and\ \citenamefont {Wolff}}]{NV_review}%
  \BibitemOpen
  \bibfield  {author} {\bibinfo {author} {\bibfnamefont {C.}~\bibnamefont
  {Tserkezis}}, \bibinfo {author} {\bibfnamefont {A.~I.}\ \bibnamefont
  {Fern{\'{a}}ndez-Dom{\'{\i}}nguez}}, \bibinfo {author} {\bibfnamefont
  {P.~A.~D.}\ \bibnamefont {Gon{\c{c}}alves}}, \bibinfo {author} {\bibfnamefont
  {F.}~\bibnamefont {Todisco}}, \bibinfo {author} {\bibfnamefont {J.~D.}\
  \bibnamefont {Cox}}, \bibinfo {author} {\bibfnamefont {K.}~\bibnamefont
  {Busch}}, \bibinfo {author} {\bibfnamefont {N.}~\bibnamefont {Stenger}},
  \bibinfo {author} {\bibfnamefont {S.~I.}\ \bibnamefont {Bozhevolnyi}},
  \bibinfo {author} {\bibfnamefont {N.~A.}\ \bibnamefont {Mortensen}},\ and\
  \bibinfo {author} {\bibfnamefont {C.}~\bibnamefont {Wolff}},\ }\bibfield
  {title} {\bibinfo {title} {On the applicability of quantum-optical concepts
  in strong-coupling nanophotonics},\ }\href
  {https://doi.org/10.1088/1361-6633/aba348} {\bibfield  {journal} {\bibinfo
  {journal} {Reports on Progress in Physics}\ }\textbf {\bibinfo {volume}
  {83}},\ \bibinfo {pages} {082401} (\bibinfo {year} {2020})}\BibitemShut
  {NoStop}%
\bibitem [{\citenamefont {Kongsuwan}\ \emph {et~al.}(2020)\citenamefont
  {Kongsuwan}, \citenamefont {Demetriadou}, \citenamefont {Horton},
  \citenamefont {Chikkaraddy}, \citenamefont {Baumberg},\ and\ \citenamefont
  {Hess}}]{Nuttawut_quasimodes}%
  \BibitemOpen
  \bibfield  {author} {\bibinfo {author} {\bibfnamefont {N.}~\bibnamefont
  {Kongsuwan}}, \bibinfo {author} {\bibfnamefont {A.}~\bibnamefont
  {Demetriadou}}, \bibinfo {author} {\bibfnamefont {M.}~\bibnamefont {Horton}},
  \bibinfo {author} {\bibfnamefont {R.}~\bibnamefont {Chikkaraddy}}, \bibinfo
  {author} {\bibfnamefont {J.~J.}\ \bibnamefont {Baumberg}},\ and\ \bibinfo
  {author} {\bibfnamefont {O.}~\bibnamefont {Hess}},\ }\bibfield  {title}
  {\bibinfo {title} {Plasmonic nanocavity modes: From near-field to far-field
  radiation},\ }\href {https://doi.org/10.1021/acsphotonics.9b01445} {\bibfield
   {journal} {\bibinfo  {journal} {ACS Photonics}\ }\textbf {\bibinfo {volume}
  {7}},\ \bibinfo {pages} {463} (\bibinfo {year} {2020})},\ \Eprint
  {https://arxiv.org/abs/https://doi.org/10.1021/acsphotonics.9b01445}
  {https://doi.org/10.1021/acsphotonics.9b01445} \BibitemShut {NoStop}%
\bibitem [{\citenamefont {Franke}\ \emph {et~al.}(2019)\citenamefont {Franke},
  \citenamefont {Hughes}, \citenamefont {Dezfouli}, \citenamefont {Kristensen},
  \citenamefont {Busch}, \citenamefont {Knorr},\ and\ \citenamefont
  {Richter}}]{Hughes_quasimodes}%
  \BibitemOpen
  \bibfield  {author} {\bibinfo {author} {\bibfnamefont {S.}~\bibnamefont
  {Franke}}, \bibinfo {author} {\bibfnamefont {S.}~\bibnamefont {Hughes}},
  \bibinfo {author} {\bibfnamefont {M.~K.}\ \bibnamefont {Dezfouli}}, \bibinfo
  {author} {\bibfnamefont {P.~T.}\ \bibnamefont {Kristensen}}, \bibinfo
  {author} {\bibfnamefont {K.}~\bibnamefont {Busch}}, \bibinfo {author}
  {\bibfnamefont {A.}~\bibnamefont {Knorr}},\ and\ \bibinfo {author}
  {\bibfnamefont {M.}~\bibnamefont {Richter}},\ }\bibfield  {title} {\bibinfo
  {title} {Quantization of quasinormal modes for open cavities and plasmonic
  cavity quantum electrodynamics},\ }\href
  {https://doi.org/10.1103/PhysRevLett.122.213901} {\bibfield  {journal}
  {\bibinfo  {journal} {Phys. Rev. Lett.}\ }\textbf {\bibinfo {volume} {122}},\
  \bibinfo {pages} {213901} (\bibinfo {year} {2019})}\BibitemShut {NoStop}%
\bibitem [{\citenamefont {Breuer}\ and\ \citenamefont
  {Petruccione}(2002)}]{Breuer}%
  \BibitemOpen
  \bibfield  {author} {\bibinfo {author} {\bibfnamefont {H.~P.}\ \bibnamefont
  {Breuer}}\ and\ \bibinfo {author} {\bibfnamefont {F.}~\bibnamefont
  {Petruccione}},\ }\href@noop {} {\emph {\bibinfo {title} {The theory of open
  quantum systems}}}\ (\bibinfo  {publisher} {Oxford University Press},\
  \bibinfo {address} {Great Clarendon Street},\ \bibinfo {year}
  {2002})\BibitemShut {NoStop}%
\bibitem [{\citenamefont {Walls}\ and\ \citenamefont
  {Milburn}(2007)}]{walls2007quantum}%
  \BibitemOpen
  \bibfield  {author} {\bibinfo {author} {\bibfnamefont {D.~F.}\ \bibnamefont
  {Walls}}\ and\ \bibinfo {author} {\bibfnamefont {G.~J.}\ \bibnamefont
  {Milburn}},\ }\href@noop {} {\emph {\bibinfo {title} {Quantum optics}}}\
  (\bibinfo  {publisher} {Springer Science \& Business Media},\ \bibinfo {year}
  {2007})\BibitemShut {NoStop}%
\bibitem [{\citenamefont {Laussy}\ \emph
  {et~al.}(2009{\natexlab{a}})\citenamefont {Laussy}, \citenamefont {del
  Valle},\ and\ \citenamefont {Tejedor}}]{ValleI}%
  \BibitemOpen
  \bibfield  {author} {\bibinfo {author} {\bibfnamefont {F.~P.}\ \bibnamefont
  {Laussy}}, \bibinfo {author} {\bibfnamefont {E.}~\bibnamefont {del Valle}},\
  and\ \bibinfo {author} {\bibfnamefont {C.}~\bibnamefont {Tejedor}},\
  }\bibfield  {title} {\bibinfo {title} {Luminescence spectra of quantum dots
  in microcavities. {I}. {Bosons}},\ }\href
  {https://doi.org/10.1103/PhysRevB.79.235325} {\bibfield  {journal} {\bibinfo
  {journal} {Physical Review B}\ }\textbf {\bibinfo {volume} {79}},\ \bibinfo
  {pages} {235325} (\bibinfo {year} {2009}{\natexlab{a}})},\ \bibinfo {note}
  {publisher: American Physical Society}\BibitemShut {NoStop}%
\bibitem [{\citenamefont {del Valle}\ \emph {et~al.}(2009)\citenamefont {del
  Valle}, \citenamefont {Laussy},\ and\ \citenamefont {Tejedor}}]{ValleII}%
  \BibitemOpen
  \bibfield  {author} {\bibinfo {author} {\bibfnamefont {E.}~\bibnamefont {del
  Valle}}, \bibinfo {author} {\bibfnamefont {F.~P.}\ \bibnamefont {Laussy}},\
  and\ \bibinfo {author} {\bibfnamefont {C.}~\bibnamefont {Tejedor}},\
  }\bibfield  {title} {\bibinfo {title} {Luminescence spectra of quantum dots
  in microcavities. {II}. {Fermions}},\ }\href
  {https://doi.org/10.1103/PhysRevB.79.235326} {\bibfield  {journal} {\bibinfo
  {journal} {Physical Review B}\ }\textbf {\bibinfo {volume} {79}},\ \bibinfo
  {pages} {235326} (\bibinfo {year} {2009})},\ \bibinfo {note} {publisher:
  American Physical Society}\BibitemShut {NoStop}%
\bibitem [{\citenamefont {Zengin}\ \emph {et~al.}(2015)\citenamefont {Zengin},
  \citenamefont {Wers\"{a}ll}, \citenamefont {Nilsson}, \citenamefont
  {Antosiewicz}, \citenamefont {K\"{a}ll},\ and\ \citenamefont
  {Shegai}}]{Zenging}%
  \BibitemOpen
  \bibfield  {author} {\bibinfo {author} {\bibfnamefont {G.}~\bibnamefont
  {Zengin}}, \bibinfo {author} {\bibfnamefont {M.}~\bibnamefont {Wers\"{a}ll}},
  \bibinfo {author} {\bibfnamefont {S.}~\bibnamefont {Nilsson}}, \bibinfo
  {author} {\bibfnamefont {T.~J.}\ \bibnamefont {Antosiewicz}}, \bibinfo
  {author} {\bibfnamefont {M.}~\bibnamefont {K\"{a}ll}},\ and\ \bibinfo
  {author} {\bibfnamefont {T.}~\bibnamefont {Shegai}},\ }\bibfield  {title}
  {\bibinfo {title} {Realizing {Strong} {Light}-{Matter} {Interactions} between
  {Single}-{Nanoparticle} {Plasmons} and {Molecular} {Excitons} at {Ambient}
  {Conditions}},\ }\href {https://doi.org/10.1103/PhysRevLett.114.157401}
  {\bibfield  {journal} {\bibinfo  {journal} {Physical Review Letters}\
  }\textbf {\bibinfo {volume} {114}},\ \bibinfo {pages} {157401} (\bibinfo
  {year} {2015})},\ \bibinfo {note} {publisher: American Physical
  Society}\BibitemShut {NoStop}%
\bibitem [{\citenamefont {Sugawara}\ \emph {et~al.}(2006)\citenamefont
  {Sugawara}, \citenamefont {Kelf}, \citenamefont {Baumberg}, \citenamefont
  {Abdelsalam},\ and\ \citenamefont {Bartlett}}]{Sugawara}%
  \BibitemOpen
  \bibfield  {author} {\bibinfo {author} {\bibfnamefont {Y.}~\bibnamefont
  {Sugawara}}, \bibinfo {author} {\bibfnamefont {T.~A.}\ \bibnamefont {Kelf}},
  \bibinfo {author} {\bibfnamefont {J.~J.}\ \bibnamefont {Baumberg}}, \bibinfo
  {author} {\bibfnamefont {M.~E.}\ \bibnamefont {Abdelsalam}},\ and\ \bibinfo
  {author} {\bibfnamefont {P.~N.}\ \bibnamefont {Bartlett}},\ }\bibfield
  {title} {\bibinfo {title} {Strong {Coupling} between {Localized} {Plasmons}
  and {Organic} {Excitons} in {Metal} {Nanovoids}},\ }\href
  {https://doi.org/10.1103/PhysRevLett.97.266808} {\bibfield  {journal}
  {\bibinfo  {journal} {Physical Review Letters}\ }\textbf {\bibinfo {volume}
  {97}},\ \bibinfo {pages} {266808} (\bibinfo {year} {2006})},\ \bibinfo {note}
  {publisher: American Physical Society}\BibitemShut {NoStop}%
\bibitem [{\citenamefont {Schlather}\ \emph {et~al.}(2013)\citenamefont
  {Schlather}, \citenamefont {Large}, \citenamefont {Urban}, \citenamefont
  {Nordlander},\ and\ \citenamefont {Halas}}]{Schlatner}%
  \BibitemOpen
  \bibfield  {author} {\bibinfo {author} {\bibfnamefont {A.~E.}\ \bibnamefont
  {Schlather}}, \bibinfo {author} {\bibfnamefont {N.}~\bibnamefont {Large}},
  \bibinfo {author} {\bibfnamefont {A.~S.}\ \bibnamefont {Urban}}, \bibinfo
  {author} {\bibfnamefont {P.}~\bibnamefont {Nordlander}},\ and\ \bibinfo
  {author} {\bibfnamefont {N.~J.}\ \bibnamefont {Halas}},\ }\bibfield  {title}
  {\bibinfo {title} {Near-{Field} {Mediated} {Plexcitonic} {Coupling} and
  {Giant} {Rabi} {Splitting} in {Individual} {Metallic} {Dimers}},\ }\href
  {https://doi.org/10.1021/nl4014887} {\bibfield  {journal} {\bibinfo
  {journal} {Nano Letters}\ }\textbf {\bibinfo {volume} {13}},\ \bibinfo
  {pages} {3281} (\bibinfo {year} {2013})},\ \bibinfo {note} {publisher:
  American Chemical Society}\BibitemShut {NoStop}%
\bibitem [{\citenamefont {Brune}\ \emph {et~al.}(1992)\citenamefont {Brune},
  \citenamefont {Haroche}, \citenamefont {Raimond}, \citenamefont
  {Davidovich},\ and\ \citenamefont {Zagury}}]{Haroche}%
  \BibitemOpen
  \bibfield  {author} {\bibinfo {author} {\bibfnamefont {M.}~\bibnamefont
  {Brune}}, \bibinfo {author} {\bibfnamefont {S.}~\bibnamefont {Haroche}},
  \bibinfo {author} {\bibfnamefont {J.~M.}\ \bibnamefont {Raimond}}, \bibinfo
  {author} {\bibfnamefont {L.}~\bibnamefont {Davidovich}},\ and\ \bibinfo
  {author} {\bibfnamefont {N.}~\bibnamefont {Zagury}},\ }\bibfield  {title}
  {\bibinfo {title} {Manipulation of photons in a cavity by dispersive
  atom-field coupling: {Quantum}-nondemolition measurements and generation of
  ``schr\"{o}dinger cat'' states},\ }\href
  {https://doi.org/10.1103/PhysRevA.45.5193} {\bibfield  {journal} {\bibinfo
  {journal} {Physical Review A}\ }\textbf {\bibinfo {volume} {45}},\ \bibinfo
  {pages} {5193} (\bibinfo {year} {1992})},\ \bibinfo {note} {publisher:
  American Physical Society}\BibitemShut {NoStop}%
\bibitem [{\citenamefont {Law}\ and\ \citenamefont {Eberly}(1996)}]{EberlyLaw}%
  \BibitemOpen
  \bibfield  {author} {\bibinfo {author} {\bibfnamefont {C.~K.}\ \bibnamefont
  {Law}}\ and\ \bibinfo {author} {\bibfnamefont {J.~H.}\ \bibnamefont
  {Eberly}},\ }\bibfield  {title} {\bibinfo {title} {Arbitrary {Control} of a
  {Quantum} {Electromagnetic} {Field}},\ }\href
  {https://doi.org/10.1103/PhysRevLett.76.1055} {\bibfield  {journal} {\bibinfo
   {journal} {Physical Review Letters}\ }\textbf {\bibinfo {volume} {76}},\
  \bibinfo {pages} {1055} (\bibinfo {year} {1996})},\ \bibinfo {note}
  {publisher: American Physical Society}\BibitemShut {NoStop}%
\bibitem [{\citenamefont {Laussy}\ \emph
  {et~al.}(2009{\natexlab{b}})\citenamefont {Laussy}, \citenamefont {del
  Valle},\ and\ \citenamefont {Tejedor}}]{laussy_luminescence_2009}%
  \BibitemOpen
  \bibfield  {author} {\bibinfo {author} {\bibfnamefont {F.~P.}\ \bibnamefont
  {Laussy}}, \bibinfo {author} {\bibfnamefont {E.}~\bibnamefont {del Valle}},\
  and\ \bibinfo {author} {\bibfnamefont {C.}~\bibnamefont {Tejedor}},\
  }\bibfield  {title} {\bibinfo {title} {Luminescence spectra of quantum dots
  in microcavities. {I}. {Bosons}},\ }\href
  {https://doi.org/10.1103/PhysRevB.79.235325} {\bibfield  {journal} {\bibinfo
  {journal} {Physical Review B}\ }\textbf {\bibinfo {volume} {79}},\ \bibinfo
  {pages} {235325} (\bibinfo {year} {2009}{\natexlab{b}})},\ \bibinfo {note}
  {publisher: American Physical Society}\BibitemShut {NoStop}%
\bibitem [{\citenamefont {Wineland}(2013)}]{Wineland_lecture}%
  \BibitemOpen
  \bibfield  {author} {\bibinfo {author} {\bibfnamefont {D.~J.}\ \bibnamefont
  {Wineland}},\ }\bibfield  {title} {\bibinfo {title} {Nobel lecture:
  Superposition, entanglement, and raising schr\"odinger's cat},\ }\href
  {https://doi.org/10.1103/RevModPhys.85.1103} {\bibfield  {journal} {\bibinfo
  {journal} {Rev. Mod. Phys.}\ }\textbf {\bibinfo {volume} {85}},\ \bibinfo
  {pages} {1103} (\bibinfo {year} {2013})}\BibitemShut {NoStop}%
\bibitem [{\citenamefont {Baranov}\ \emph {et~al.}(2017)\citenamefont
  {Baranov}, \citenamefont {Wersäll}, \citenamefont {Cuadra}, \citenamefont
  {Antosiewicz},\ and\ \citenamefont {Shegai}}]{baranov_novel_2017}%
  \BibitemOpen
  \bibfield  {author} {\bibinfo {author} {\bibfnamefont {D.~G.}\ \bibnamefont
  {Baranov}}, \bibinfo {author} {\bibfnamefont {M.}~\bibnamefont {Wersäll}},
  \bibinfo {author} {\bibfnamefont {J.}~\bibnamefont {Cuadra}}, \bibinfo
  {author} {\bibfnamefont {T.~J.}\ \bibnamefont {Antosiewicz}},\ and\ \bibinfo
  {author} {\bibfnamefont {T.}~\bibnamefont {Shegai}},\ }\bibfield  {title}
  {\bibinfo {title} {Novel {Nanostructures} and {Materials} for {Strong}
  {Light}–{Matter} {Interactions}},\ }\href
  {https://doi.org/10.1021/acsphotonics.7b00674} {\bibfield  {journal}
  {\bibinfo  {journal} {ACS Photonics}\ }\textbf {\bibinfo {volume} {5}},\
  \bibinfo {pages} {24} (\bibinfo {year} {2017})}\BibitemShut {NoStop}%
\bibitem [{\citenamefont {Huang}\ \emph {et~al.}(2020)\citenamefont {Huang},
  \citenamefont {Wu},\ and\ \citenamefont {Yu}}]{huang_rabi_2020}%
  \BibitemOpen
  \bibfield  {author} {\bibinfo {author} {\bibfnamefont {Y.}~\bibnamefont
  {Huang}}, \bibinfo {author} {\bibfnamefont {F.}~\bibnamefont {Wu}},\ and\
  \bibinfo {author} {\bibfnamefont {L.}~\bibnamefont {Yu}},\ }\bibfield
  {title} {\bibinfo {title} {Rabi oscillation study of strong coupling in a
  plasmonic nanocavity},\ }\bibfield  {journal} {\bibinfo  {journal} {New
  Journal of Physics}\ }\textbf {\bibinfo {volume} {22}},\ \href
  {https://doi.org/10.1088/1367-2630/ab9222} {10.1088/1367-2630/ab9222}
  (\bibinfo {year} {2020})\BibitemShut {NoStop}%
\end{thebibliography}%

\appendix
	\section{Auxiliary models}\label{auxmodels}
	Here we briefly list the Hamiltonians and Lindblad operators for the auxiliary models used throughout the paper. In all cases the field mode is described by the same annihilation operator $\hat a$ (and corresponding Hilbert space) as in the main text. In turn, photon loss and coherent cavity driving are described by the same operators shown in the main text (Section~\ref{MEQsec}).
	What changes between the different models is the structure of the emitter, and the associated Lindblad operators.

\subsection{Jaynes-Cummings Model} 
The Hamiltonian of the JC model adopted in the main text reads
	\begin{align}
	    H_\mathrm{JC}&=\omega_0\left(\hat a^\dagger \hat a+\proj{e}\right)+g_\mathrm{eff}(\hat a^\dagger \kebra{g}{e}+ \hat a\kebra{e}{g}),\label{JC-H}
	\end{align}
where the emitter is a two-level system with excited state $\ket e$ and ground state $\ket g$. The Lindblad operators for emitter decay, dephasing and incoherent pumping are respectively
	\begin{align}
	    L&=\sqrt{\gamma}\kebra{g}{e},\\ L&=\sqrt{\gamma_d}\proj{e},\\
	    L&=\sqrt{\Lambda}\kebra{e}{g}
	\end{align}

\subsection{Tavis-Cummings model with \texorpdfstring{$N$}{N} atoms}
The TC model used in the main text, with $N$ atoms and effective coupling $g_\mathrm{eff}$, reads
	\begin{align}
H_\mathrm{TC}&=\omega_0\hat a^\dagger\hat a +\omega_0 \tfrac{\hat J_z}{2}+\frac{g_\mathrm{eff}}{\sqrt{N}}(\hat J^+\hat a+\hat a^\dagger\hat J^-),\label{angolone}
\end{align}
where $\hat J_x,\hat J_y,\hat J_z$ are spin-$N/2$ operators, and $\hat J^{\pm}=\hat J_x\pm i\hat J_y$ are the spin-ladder operators. The Lindblad operators for emitter decay, dephasing and incoherent pumping are respectively
\begin{align}
  L&=\sqrt{\gamma}\hat{J}^-,\\ L&=\sqrt{\gamma_d}\hat{J}_z/2,\\ L&=\sqrt{\Lambda}\hat{J}^+
\end{align}
\subsection{Coupled oscillators model}
In the rotating wave approximation, the Hamiltonian of two coupled harmonic oscillators (both having the same frequency) reads
	\begin{align}
	    H_\mathrm{osc}&=\omega_0\left(\hat a^\dagger \hat a+\hat b^\dagger \hat b\right)+g_\mathrm{eff}(\hat a^\dagger \hat b+\hat b^\dagger \hat a),\label{two-osc}
	\end{align}
	where $\hat a,\hat b$ are  annihilation operators of distinct Bosonic modes. The Lindblad operator for emitter decay, dephasing and incoherent pumping are respectively
	\begin{align}
	    L&=\sqrt{\gamma}\hat b,\\  L&=\sqrt{\gamma_d}\hat b^\dagger \hat b,\\ L&=\sqrt{\Lambda}\hat b^\dagger.
	\end{align}
\end{document}